\documentclass[12pt]{article}
\usepackage{epsfig,amssymb,amsmath,psfrag}
\usepackage{cite}

\def\bigsk{\bigskip}
\def\cyc{{\rm cyc}} 
\def\rev{{\rm rev}} 
\def\noin{}
\def\aa{m}
\def\Aa{a}
\def\Ab{b}
\def\Ac{c}
\def\Ad{d}
\def\Ae{e}
\def\Af{0}
\def\Ag{f}
\def\Ah{g}
\def\Ai{0}
\def\Aj{h}
\def\Ak{0}
\def\Al{i}
\def\Am{0}
\def\An{j}
\def\setal{ \{ \alpha \} }
\def\setbe{ \{ \beta \} }
\def\setbeT{ \{ \beta^T \} }

\def\plus{ ~{\oplus \atop +}~}
\def\equals{ ~{= \atop = }~}
\def\nicefrac#1#2{\textstyle{\frac{#1}{#2}}}

\def\Mila{ M_{i\la} }
\def\Ala{ A_\la }

\def\vla{ {v_\la} }
\def\rep#1#2{ v \left(  ~#1  ~ | ~ #2 ~\right) }
\def\repl#1#2{ \vla \left(  ~#1  ~ | ~ #2 ~\right) }
\def\repi#1#2{ v_i \left(  ~#1  ~ | ~ #2 ~\right) }
\def\repli#1#2{ \vla_i \left(  ~#1  ~ | ~ #2 ~\right) }
\def\repij#1#2#3{ v_i \left(  ~#1  ~ | ~ #2\: ,  #3 ~\right) }
\def\replij#1#2#3{ \vla_i \left(  ~#1  ~ | ~ #2\: ,  #3 ~\right) }

\def\ind{ {\rm (ind)} }
\def\reg{ {\rm reg} }
\def\ncon{n_{\rm null}}
\def\nsing{n_{\rm s}}
\def\ndoub{n_{\rm d}}
\def\ntrip{n_{\rm t}}

\def\f{\tilde{f}}

\def\Ell{{(L)}}

\def\Elltk{{(L,2k)}}
\def\Elltkp{{(L,2k+1)}}

\def\Tete{{(2\ell,2 \ell)}}
\def\Tetemo{{(2\ell,2 \ell-1)}}
\def\Tetemt{{(2\ell,2 \ell-2)}}
\def\Teptep{{(2\ell+1,2 \ell+1)}}
\def\Tepte{{(2\ell+1,2 \ell)}}
\def\Teptemo{{(2\ell+1,2 \ell-1)}}
\def\Teptemt{{(2\ell+1,2 \ell-2)}}

\def\Ellplus{{(L+1)}}
\def\Zero{{(0)}}
\def\Zerozero{{(0,0)}}
\def\One{{(1)}}
\def\Onezero{{(1,0)}}
\def\Oneone{{(1,1)}}
\def\Two{{(2)}}
\def\Twozero{{(2,0)}}
\def\Twoone{{(2,1)}}
\def\Twotwo{{(2,2)}}

\def\cA {  {\cal A}  }
\def\la {\lambda}
\def\sig {\sigma}
\def\ka {\kappa}

\def \be  {\begin{equation}}
\def \ee  {\end{equation}}
\def \ba  {\begin{eqnarray}}
\def \ea  {\end{eqnarray}}

\newcommand{\nn}{\nonumber}

\def \Tr {\mathop{\rm Tr}\nolimits}

\def\half{\textstyle{1 \over 2}}
\def\tenth{\textstyle{1 \over 10}}
\def\third{\textstyle{1 \over 3}}
\def\twothirds{\textstyle{2 \over 3}}
\def\fourthirds{\textstyle{4 \over 3}}
\def\fourth{\textstyle{1 \over 4}}

\def\eqn#1{eq.~(\ref{#1})} \def\Eqn#1{Equation~(\ref{#1})}
\def\eqns#1#2{eqs.~(\ref{#1}) and~(\ref{#2})}

\font\cmss=cmss10 at 11pt \font\cmsss=cmss8 at 8pt

\def\mininbar{\vrule height.75ex width.3pt depth0pt}
\def\cc{\relax\,\hbox{$\mininbar\kern-.2em{\hbox{\rm\tiny C}}$}}
\def\IZ{\relax\ifmmode\mathchoice
{\hbox{\cmss Z\kern-.4em Z}}{\hbox{\cmss Z\kern-.4em Z}}
{\lower.4pt\hbox{\cmsss Z\kern-.4em Z}}
{\lower1.2pt\hbox{\cmsss Z\kern-.4em Z}}\else{\cmss Z\kern-.4em Z}\fi}

\def\one{  {\vcenter  {\vbox  
              {\hrule height.4pt
               \hbox {\vrule width.4pt  height3pt  
                      \kern3pt 
                      \vrule width.4pt  height3pt }
               \hrule height.4pt}
                         }
                   }
           }
\def\two{  {\vcenter  {\vbox  
              {\hrule height.4pt
               \hbox {\vrule width.4pt  height3pt  
                      \kern3pt 
                      \vrule width.4pt  height3pt 
                      \kern3pt
                      \vrule width.4pt height3pt}
               \hrule height.4pt}
                         }
              }
           }
\def\three{  {\vcenter  {\vbox  
              {\hrule height.4pt
               \hbox {\vrule width.4pt  height3pt  
                      \kern3pt 
                      \vrule width.4pt  height3pt 
                      \kern3pt
                      \vrule width.4pt  height3pt 
                      \kern3pt
                      \vrule width.4pt height3pt}
               \hrule height.4pt}
                         }
              }
           }
\def\four{  {\vcenter  {\vbox  
              {\hrule height.4pt
               \hbox {\vrule width.4pt  height3pt  
                      \kern3pt 
                      \vrule width.4pt  height3pt 
                      \kern3pt
                      \vrule width.4pt  height3pt 
                      \kern3pt
                      \vrule width.4pt  height3pt 
                      \kern3pt
                      \vrule width.4pt height3pt}
               \hrule height.4pt}
                         }
              }
           }
\def\five{  {\vcenter  {\vbox  
              {\hrule height.4pt
               \hbox {\vrule width.4pt  height3pt  
                      \kern3pt 
                      \vrule width.4pt  height3pt 
                      \kern3pt 
                      \vrule width.4pt  height3pt 
                      \kern3pt
                      \vrule width.4pt  height3pt 
                      \kern3pt
                      \vrule width.4pt  height3pt 
                      \kern3pt
                      \vrule width.4pt height3pt}
               \hrule height.4pt}
                         }
              }
           }
\def\six{  {\vcenter  {\vbox  
              {\hrule height.4pt
               \hbox {\vrule width.4pt  height3pt  
                      \kern3pt 
                      \vrule width.4pt  height3pt 
                      \kern3pt 
                      \vrule width.4pt  height3pt 
                      \kern3pt 
                      \vrule width.4pt  height3pt 
                      \kern3pt
                      \vrule width.4pt  height3pt 
                      \kern3pt
                      \vrule width.4pt  height3pt 
                      \kern3pt
                      \vrule width.4pt height3pt}
               \hrule height.4pt}
                         }
              }
           }
\def\oneone{ {\vcenter  {\vbox  
              {\hrule height.4pt
               \hbox {\vrule width.4pt  height3pt  
                      \kern3pt 
                      \vrule width.4pt  height3pt }
               \hrule height.4pt
               \hbox {\vrule width.4pt  height3pt  
                      \kern3pt 
                      \vrule width.4pt  height3pt }
               \hrule height.4pt}
                         }
              }
           }
\def\twoone{ 
              {\vcenter  {\vbox  
              {\hrule height.4pt
               \hbox {\vrule width.4pt  height3pt  
                      \kern3pt 
                      \vrule width.4pt  height3pt 
                      \kern3pt
                      \vrule width.4pt height3pt}
               \hrule height.4pt
               \hbox {\vrule width.4pt  height3pt  
                      \kern3pt 
                      \vrule width.4pt  height3pt }
               \hrule height.4pt width3.8pt}
                         }
              }
           }
\def\twotwo{ 
              {\vcenter  {\vbox  
              {\hrule height.4pt
               \hbox {\vrule width.4pt  height3pt  
                      \kern3pt 
                      \vrule width.4pt  height3pt 
                      \kern3pt
                      \vrule width.4pt height3pt}
               \hrule height.4pt
               \hbox {\vrule width.4pt  height3pt  
                      \kern3pt 
                      \vrule width.4pt  height3pt 
                      \kern3pt
                      \vrule width.4pt height3pt}
               \hrule height.4pt}
                         }
              }
           }
\def\threeone{ 
              {\vcenter  {\vbox  
              {\hrule height.4pt
               \hbox {\vrule width.4pt  height3pt  
                      \kern3pt 
                      \vrule width.4pt  height3pt 
                      \kern3pt
                      \vrule width.4pt  height3pt 
                      \kern3pt
                      \vrule width.4pt height3pt}
               \hrule height.4pt
               \hbox {\vrule width.4pt  height3pt  
                      \kern3pt 
                      \vrule width.4pt  height3pt }
               \hrule height.4pt width3.8pt}
                         }
              }
           }
\def\threetwo{ 
              {\vcenter  {\vbox  
              {\hrule height.4pt
               \hbox {\vrule width.4pt  height3pt  
                      \kern3pt 
                      \vrule width.4pt  height3pt 
                      \kern3pt 
                      \vrule width.4pt  height3pt 
                      \kern3pt
                      \vrule width.4pt height3pt}
               \hrule height.4pt
               \hbox {\vrule width.4pt  height3pt  
                      \kern3pt 
                      \vrule width.4pt  height3pt 
                      \kern3pt
                      \vrule width.4pt height3pt}
               \hrule height.4pt width 7.2pt}
                         }
              }
           }
\def\threethree{ 
              {\vcenter  {\vbox  
              {\hrule height.4pt
               \hbox {\vrule width.4pt  height3pt  
                      \kern3pt 
                      \vrule width.4pt  height3pt 
                      \kern3pt 
                      \vrule width.4pt  height3pt 
                      \kern3pt
                      \vrule width.4pt height3pt}
               \hrule height.4pt
               \hbox {\vrule width.4pt  height3pt  
                      \kern3pt 
                      \vrule width.4pt  height3pt 
                      \kern3pt
                      \vrule width.4pt  height3pt 
                      \kern3pt
                      \vrule width.4pt height3pt}
               \hrule height.4pt }
                         }
              }
           }
\def\fourtwo{ 
              {\vcenter  {\vbox  
              {\hrule height.4pt
               \hbox {\vrule width.4pt  height3pt  
                      \kern3pt 
                      \vrule width.4pt  height3pt 
                      \kern3pt 
                      \vrule width.4pt  height3pt 
                      \kern3pt 
                      \vrule width.4pt  height3pt 
                      \kern3pt
                      \vrule width.4pt height3pt}
               \hrule height.4pt
               \hbox {\vrule width.4pt  height3pt  
                      \kern3pt 
                      \vrule width.4pt  height3pt 
                      \kern3pt
                      \vrule width.4pt height3pt}
               \hrule height.4pt width7.2pt}
                         }
              }
           }
\def\fourthree{ 
              {\vcenter  {\vbox  
              {\hrule height.4pt
               \hbox {\vrule width.4pt  height3pt  
                      \kern3pt 
                      \vrule width.4pt  height3pt 
                      \kern3pt 
                      \vrule width.4pt  height3pt 
                      \kern3pt 
                      \vrule width.4pt  height3pt 
                      \kern3pt
                      \vrule width.4pt height3pt}
               \hrule height.4pt
               \hbox {\vrule width.4pt  height3pt  
                      \kern3pt 
                      \vrule width.4pt  height3pt 
                      \kern3pt
                      \vrule width.4pt  height3pt 
                      \kern3pt
                      \vrule width.4pt height3pt}
               \hrule height.4pt width10.6pt}
                         }
              }
           }
\def\fiveone{ 
              {\vcenter  {\vbox  
              {\hrule height.4pt
               \hbox {\vrule width.4pt  height3pt  
                      \kern3pt 
                      \vrule width.4pt  height3pt 
                      \kern3pt 
                      \vrule width.4pt  height3pt 
                      \kern3pt 
                      \vrule width.4pt  height3pt 
                      \kern3pt 
                      \vrule width.4pt  height3pt 
                      \kern3pt
                      \vrule width.4pt height3pt}
               \hrule height.4pt
               \hbox {\vrule width.4pt  height3pt  
                      \kern3pt 
                      \vrule width.4pt height3pt}
               \hrule height.4pt width3.8pt}
                         }
              }
           }
\def\fivetwo{ 
              {\vcenter  {\vbox  
              {\hrule height.4pt
               \hbox {\vrule width.4pt  height3pt  
                      \kern3pt 
                      \vrule width.4pt  height3pt 
                      \kern3pt 
                      \vrule width.4pt  height3pt 
                      \kern3pt 
                      \vrule width.4pt  height3pt 
                      \kern3pt 
                      \vrule width.4pt  height3pt 
                      \kern3pt
                      \vrule width.4pt height3pt}
               \hrule height.4pt
               \hbox {\vrule width.4pt  height3pt  
                      \kern3pt 
                      \vrule width.4pt  height3pt 
                      \kern3pt
                      \vrule width.4pt height3pt}
               \hrule height.4pt width7.2pt}
                         }
              }
           }
\def\sixsix{ 
              {\vcenter  {\vbox  
              {\hrule height.4pt
               \hbox {\vrule width.4pt  height3pt  
                      \kern3pt 
                      \vrule width.4pt  height3pt 
                      \kern3pt 
                      \vrule width.4pt  height3pt 
                      \kern3pt 
                      \vrule width.4pt  height3pt 
                      \kern3pt 
                      \vrule width.4pt  height3pt 
                      \kern3pt 
                      \vrule width.4pt  height3pt 
                      \kern3pt 
                      \vrule width.4pt  height3pt }
               \hrule height.4pt
               \hbox {\vrule width.4pt  height3pt  
                      \kern3pt 
                      \vrule width.4pt  height3pt 
                      \kern3pt 
                      \vrule width.4pt  height3pt 
                      \kern3pt 
                      \vrule width.4pt  height3pt 
                      \kern3pt 
                      \vrule width.4pt  height3pt 
                      \kern3pt 
                      \vrule width.4pt  height3pt 
                      \kern3pt 
                      \vrule width.4pt  height3pt }
               \hrule height.4pt }
                         }
              }
           }
\def\oneoneone{ 
              {\vcenter  {\vbox  
              {\hrule height.4pt
               \hbox {\vrule width.4pt  height3pt  
                      \kern3pt 
                      \vrule width.4pt  height3pt }
               \hrule height.4pt
               \hbox {\vrule width.4pt  height3pt  
                      \kern3pt 
                      \vrule width.4pt  height3pt }
               \hrule height.4pt
               \hbox {\vrule width.4pt  height3pt  
                      \kern3pt 
                      \vrule width.4pt  height3pt }
               \hrule height.4pt}
                         }
              }
           }
\def\twooneone{ 
              {\vcenter  {\vbox  
              {\hrule height.4pt
               \hbox {\vrule width.4pt  height3pt  
                      \kern3pt 
                      \vrule width.4pt  height3pt 
                      \kern3pt 
                      \vrule width.4pt  height3pt }
               \hrule height.4pt
               \hbox {\vrule width.4pt  height3pt  
                      \kern3pt 
                      \vrule width.4pt  height3pt }
               \hrule height.4pt width3.8pt
               \hbox {\vrule width.4pt  height3pt  
                      \kern3pt 
                      \vrule width.4pt  height3pt }
               \hrule height.4pt width3.8pt}
                         }
              }
           }
\def\twotwoone{ 
              {\vcenter  {\vbox  
              {\hrule height.4pt
               \hbox {\vrule width.4pt  height3pt  
                      \kern3pt 
                      \vrule width.4pt  height3pt 
                      \kern3pt 
                      \vrule width.4pt  height3pt }
               \hrule height.4pt
               \hbox {\vrule width.4pt  height3pt  
                      \kern3pt 
                      \vrule width.4pt  height3pt 
                      \kern3pt 
                      \vrule width.4pt  height3pt }
               \hrule height.4pt width7.2pt
               \hbox {\vrule width.4pt  height3pt  
                      \kern3pt 
                      \vrule width.4pt  height3pt }
               \hrule height.4pt width3.8pt}
                         }
              }
           }
\def\twotwotwo{ 
              {\vcenter  {\vbox  
              {\hrule height.4pt
               \hbox {\vrule width.4pt  height3pt  
                      \kern3pt 
                      \vrule width.4pt  height3pt 
                      \kern3pt 
                      \vrule width.4pt  height3pt }
               \hrule height.4pt
               \hbox {\vrule width.4pt  height3pt  
                      \kern3pt 
                      \vrule width.4pt  height3pt 
                      \kern3pt 
                      \vrule width.4pt  height3pt }
               \hrule height.4pt
               \hbox {\vrule width.4pt  height3pt  
                      \kern3pt 
                      \vrule width.4pt  height3pt 
                      \kern3pt 
                      \vrule width.4pt  height3pt }
               \hrule height.4pt }
                         }
              }
           }
\def\threeoneone{ 
              {\vcenter  {\vbox  
              {\hrule height.4pt
               \hbox {\vrule width.4pt  height3pt  
                      \kern3pt 
                      \vrule width.4pt  height3pt 
                      \kern3pt 
                      \vrule width.4pt  height3pt 
                      \kern3pt 
                      \vrule width.4pt  height3pt }
               \hrule height.4pt
               \hbox {\vrule width.4pt  height3pt  
                      \kern3pt 
                      \vrule width.4pt  height3pt }
               \hrule height.4pt width3.8pt
               \hbox {\vrule width.4pt  height3pt  
                      \kern3pt 
                      \vrule width.4pt  height3pt }
               \hrule height.4pt width3.8pt}
                         }
              }
           }
\def\threetwoone{ 
              {\vcenter  {\vbox  
              {\hrule height.4pt
               \hbox {\vrule width.4pt  height3pt  
                      \kern3pt 
                      \vrule width.4pt  height3pt 
                      \kern3pt 
                      \vrule width.4pt  height3pt 
                      \kern3pt 
                      \vrule width.4pt  height3pt }
               \hrule height.4pt
               \hbox {\vrule width.4pt  height3pt  
                      \kern3pt 
                      \vrule width.4pt  height3pt 
                      \kern3pt 
                      \vrule width.4pt  height3pt }
               \hrule height.4pt width7.2pt
               \hbox {\vrule width.4pt  height3pt  
                      \kern3pt 
                      \vrule width.4pt  height3pt }
               \hrule height.4pt width3.8pt}
                         }
              }
           }
\def\threetwotwo{ 
              {\vcenter  {\vbox  
              {\hrule height.4pt
               \hbox {\vrule width.4pt  height3pt  
                      \kern3pt 
                      \vrule width.4pt  height3pt 
                      \kern3pt 
                      \vrule width.4pt  height3pt 
                      \kern3pt 
                      \vrule width.4pt  height3pt }
               \hrule height.4pt
               \hbox {\vrule width.4pt  height3pt  
                      \kern3pt 
                      \vrule width.4pt  height3pt 
                      \kern3pt 
                      \vrule width.4pt  height3pt }
               \hrule height.4pt width7.2pt
               \hbox {\vrule width.4pt  height3pt  
                      \kern3pt 
                      \vrule width.4pt  height3pt 
                      \kern3pt 
                      \vrule width.4pt  height3pt }
               \hrule height.4pt width7.2pt}
                         }
              }
           }
\def\threethreeone{ 
              {\vcenter  {\vbox  
              {\hrule height.4pt
               \hbox {\vrule width.4pt  height3pt  
                      \kern3pt 
                      \vrule width.4pt  height3pt 
                      \kern3pt 
                      \vrule width.4pt  height3pt 
                      \kern3pt 
                      \vrule width.4pt  height3pt }
               \hrule height.4pt
               \hbox {\vrule width.4pt  height3pt  
                      \kern3pt 
                      \vrule width.4pt  height3pt 
                      \kern3pt 
                      \vrule width.4pt  height3pt 
                      \kern3pt 
                      \vrule width.4pt  height3pt }
               \hrule height.4pt 
               \hbox {\vrule width.4pt  height3pt  
                      \kern3pt 
                      \vrule width.4pt  height3pt }
               \hrule height.4pt width3.8pt}
                         }
              }
           }
\def\fouroneone{ 
              {\vcenter  {\vbox  
              {\hrule height.4pt
               \hbox {\vrule width.4pt  height3pt  
                      \kern3pt 
                      \vrule width.4pt  height3pt 
                      \kern3pt 
                      \vrule width.4pt  height3pt 
                      \kern3pt 
                      \vrule width.4pt  height3pt 
                      \kern3pt 
                      \vrule width.4pt  height3pt }
               \hrule height.4pt
               \hbox {\vrule width.4pt  height3pt  
                      \kern3pt 
                      \vrule width.4pt  height3pt }
               \hrule height.4pt width3.8pt
               \hbox {\vrule width.4pt  height3pt  
                      \kern3pt 
                      \vrule width.4pt  height3pt }
               \hrule height.4pt width3.8pt}
                         }
              }
           }
\def\fourtwoone{ 
              {\vcenter  {\vbox  
              {\hrule height.4pt
               \hbox {\vrule width.4pt  height3pt  
                      \kern3pt 
                      \vrule width.4pt  height3pt 
                      \kern3pt 
                      \vrule width.4pt  height3pt 
                      \kern3pt 
                      \vrule width.4pt  height3pt 
                      \kern3pt 
                      \vrule width.4pt  height3pt }
               \hrule height.4pt
               \hbox {\vrule width.4pt  height3pt  
                      \kern3pt 
                      \vrule width.4pt  height3pt 
                      \kern3pt 
                      \vrule width.4pt  height3pt }
               \hrule height.4pt width7.2pt
               \hbox {\vrule width.4pt  height3pt  
                      \kern3pt 
                      \vrule width.4pt  height3pt }
               \hrule height.4pt width3.8pt}
                         }
              }
           }
\def\fiveoneone{ 
              {\vcenter  {\vbox  
              {\hrule height.4pt
               \hbox {\vrule width.4pt  height3pt  
                      \kern3pt 
                      \vrule width.4pt  height3pt 
                      \kern3pt 
                      \vrule width.4pt  height3pt 
                      \kern3pt 
                      \vrule width.4pt  height3pt 
                      \kern3pt 
                      \vrule width.4pt  height3pt 
                      \kern3pt 
                      \vrule width.4pt  height3pt }
               \hrule height.4pt
               \hbox {\vrule width.4pt  height3pt  
                      \kern3pt 
                      \vrule width.4pt  height3pt }
               \hrule height.4pt width3.8pt
               \hbox {\vrule width.4pt  height3pt  
                      \kern3pt 
                      \vrule width.4pt  height3pt }
               \hrule height.4pt width3.8pt}
                         }
              }
           }
\def\oneoneoneone{ 
              {\vcenter  {\vbox  
              {\hrule height.4pt
               \hbox {\vrule width.4pt  height3pt  
                      \kern3pt 
                      \vrule width.4pt  height3pt }
               \hrule height.4pt
               \hbox {\vrule width.4pt  height3pt  
                      \kern3pt 
                      \vrule width.4pt  height3pt }
               \hrule height.4pt
               \hbox {\vrule width.4pt  height3pt  
                      \kern3pt 
                      \vrule width.4pt  height3pt }
               \hrule height.4pt
               \hbox {\vrule width.4pt  height3pt  
                      \kern3pt 
                      \vrule width.4pt  height3pt }
               \hrule height.4pt}
                         }
              }
           }
\def\twooneoneone{ 
              {\vcenter  {\vbox  
              {\hrule height.4pt
               \hbox {\vrule width.4pt  height3pt  
                      \kern3pt 
                      \vrule width.4pt  height3pt 
                      \kern3pt 
                      \vrule width.4pt  height3pt }
               \hrule height.4pt
               \hbox {\vrule width.4pt  height3pt  
                      \kern3pt 
                      \vrule width.4pt  height3pt }
               \hrule height.4pt width 3.8pt
               \hbox {\vrule width.4pt  height3pt  
                      \kern3pt 
                      \vrule width.4pt  height3pt }
               \hrule height.4pt width3.8pt
               \hbox {\vrule width.4pt  height3pt  
                      \kern3pt 
                      \vrule width.4pt  height3pt }
               \hrule height.4pt width3.8pt}
                         }
              }
           }
\def\twotwooneone{ 
              {\vcenter  {\vbox  
              {\hrule height.4pt
               \hbox {\vrule width.4pt  height3pt  
                      \kern3pt 
                      \vrule width.4pt  height3pt 
                      \kern3pt 
                      \vrule width.4pt  height3pt }
               \hrule height.4pt
               \hbox {\vrule width.4pt  height3pt  
                      \kern3pt 
                      \vrule width.4pt  height3pt 
                      \kern3pt 
                      \vrule width.4pt  height3pt }
               \hrule height.4pt
               \hbox {\vrule width.4pt  height3pt  
                      \kern3pt 
                      \vrule width.4pt  height3pt }
               \hrule height.4pt width3.8pt
               \hbox {\vrule width.4pt  height3pt  
                      \kern3pt 
                      \vrule width.4pt  height3pt }
               \hrule height.4pt width3.8pt}
                         }
              }
           }
\def\twotwotwoone{ 
              {\vcenter  {\vbox  
              {\hrule height.4pt
               \hbox {\vrule width.4pt  height3pt  
                      \kern3pt 
                      \vrule width.4pt  height3pt 
                      \kern3pt 
                      \vrule width.4pt  height3pt }
               \hrule height.4pt
               \hbox {\vrule width.4pt  height3pt  
                      \kern3pt 
                      \vrule width.4pt  height3pt 
                      \kern3pt 
                      \vrule width.4pt  height3pt }
               \hrule height.4pt
               \hbox {\vrule width.4pt  height3pt  
                      \kern3pt 
                      \vrule width.4pt  height3pt 
                      \kern3pt 
                      \vrule width.4pt  height3pt }
               \hrule height.4pt
               \hbox {\vrule width.4pt  height3pt  
                      \kern3pt 
                      \vrule width.4pt  height3pt }
               \hrule height.4pt width3.8pt}
                         }
              }
           }
\def\threeoneoneone{ 
              {\vcenter  {\vbox  
              {\hrule height.4pt
               \hbox {\vrule width.4pt  height3pt  
                      \kern3pt 
                      \vrule width.4pt  height3pt 
                      \kern3pt 
                      \vrule width.4pt  height3pt 
                      \kern3pt 
                      \vrule width.4pt  height3pt }
               \hrule height.4pt
               \hbox {\vrule width.4pt  height3pt  
                      \kern3pt 
                      \vrule width.4pt  height3pt }
               \hrule height.4pt width3.8pt
               \hbox {\vrule width.4pt  height3pt  
                      \kern3pt 
                      \vrule width.4pt  height3pt }
               \hrule height.4pt width3.8pt
               \hbox {\vrule width.4pt  height3pt  
                      \kern3pt 
                      \vrule width.4pt  height3pt }
               \hrule height.4pt width3.8pt}
                         }
              }
           }
\def\threetwooneone{ 
              {\vcenter  {\vbox  
              {\hrule height.4pt
               \hbox {\vrule width.4pt  height3pt  
                      \kern3pt 
                      \vrule width.4pt  height3pt 
                      \kern3pt 
                      \vrule width.4pt  height3pt 
                      \kern3pt 
                      \vrule width.4pt  height3pt }
               \hrule height.4pt
               \hbox {\vrule width.4pt  height3pt  
                      \kern3pt 
                      \vrule width.4pt  height3pt 
                      \kern3pt 
                      \vrule width.4pt  height3pt }
               \hrule height.4pt width7.2pt
               \hbox {\vrule width.4pt  height3pt  
                      \kern3pt 
                      \vrule width.4pt  height3pt }
               \hrule height.4pt width3.8pt
               \hbox {\vrule width.4pt  height3pt  
                      \kern3pt 
                      \vrule width.4pt  height3pt }
               \hrule height.4pt width3.8pt}
                         }
              }
           }
\def\fouroneoneone{ 
              {\vcenter  {\vbox  
              {\hrule height.4pt
               \hbox {\vrule width.4pt  height3pt  
                      \kern3pt 
                      \vrule width.4pt  height3pt 
                      \kern3pt 
                      \vrule width.4pt  height3pt 
                      \kern3pt 
                      \vrule width.4pt  height3pt 
                      \kern3pt 
                      \vrule width.4pt  height3pt }
               \hrule height.4pt
               \hbox {\vrule width.4pt  height3pt  
                      \kern3pt 
                      \vrule width.4pt  height3pt }
               \hrule height.4pt width3.8pt
               \hbox {\vrule width.4pt  height3pt  
                      \kern3pt 
                      \vrule width.4pt  height3pt }
               \hrule height.4pt width3.8pt
               \hbox {\vrule width.4pt  height3pt  
                      \kern3pt 
                      \vrule width.4pt  height3pt }
               \hrule height.4pt width3.8pt}
                         }
              }
           }
\def\sixsixtwoone{ 
              {\vcenter  {\vbox  
              {\hrule height.4pt
               \hbox {\vrule width.4pt  height3pt  
                      \kern3pt 
                      \vrule width.4pt  height3pt 
                      \kern3pt 
                      \vrule width.4pt  height3pt 
                      \kern3pt 
                      \vrule width.4pt  height3pt 
                      \kern3pt 
                      \vrule width.4pt  height3pt 
                      \kern3pt 
                      \vrule width.4pt  height3pt 
                      \kern3pt 
                      \vrule width.4pt  height3pt }
               \hrule height.4pt
               \hbox {\vrule width.4pt  height3pt  
                      \kern3pt 
                      \vrule width.4pt  height3pt 
                      \kern3pt 
                      \vrule width.4pt  height3pt 
                      \kern3pt 
                      \vrule width.4pt  height3pt 
                      \kern3pt 
                      \vrule width.4pt  height3pt 
                      \kern3pt 
                      \vrule width.4pt  height3pt 
                      \kern3pt 
                      \vrule width.4pt  height3pt }
               \hrule height.4pt
               \hbox {\vrule width.4pt  height3pt  
                      \kern3pt 
                      \vrule width.4pt  height3pt 
                      \kern3pt 
                      \vrule width.4pt  height3pt }
               \hrule height.4pt width7.2pt
               \hbox {\vrule width.4pt  height3pt  
                      \kern3pt 
                      \vrule width.4pt  height3pt }
               \hrule height.4pt width3.8pt}
                         }
              }
           }
\def\oneoneoneoneone{ 
              {\vcenter  {\vbox  
              {\hrule height.4pt
               \hbox {\vrule width.4pt  height3pt  
                      \kern3pt 
                      \vrule width.4pt  height3pt }
               \hrule height.4pt
               \hbox {\vrule width.4pt  height3pt  
                      \kern3pt 
                      \vrule width.4pt  height3pt }
               \hrule height.4pt width3.8pt
               \hbox {\vrule width.4pt  height3pt  
                      \kern3pt 
                      \vrule width.4pt  height3pt }
               \hrule height.4pt width3.8pt
               \hbox {\vrule width.4pt  height3pt  
                      \kern3pt 
                      \vrule width.4pt  height3pt }
               \hrule height.4pt width3.8pt
               \hbox {\vrule width.4pt  height3pt  
                      \kern3pt 
                      \vrule width.4pt  height3pt }
               \hrule height.4pt width3.8pt}
                         }
              }
           }
\def\twooneoneoneone{ 
              {\vcenter  {\vbox  
              {\hrule height.4pt
               \hbox {\vrule width.4pt  height3pt  
                      \kern3pt 
                      \vrule width.4pt  height3pt 
                      \kern3pt 
                      \vrule width.4pt  height3pt }
               \hrule height.4pt
               \hbox {\vrule width.4pt  height3pt  
                      \kern3pt 
                      \vrule width.4pt  height3pt }
               \hrule height.4pt width3.8pt
               \hbox {\vrule width.4pt  height3pt  
                      \kern3pt 
                      \vrule width.4pt  height3pt }
               \hrule height.4pt width3.8pt
               \hbox {\vrule width.4pt  height3pt  
                      \kern3pt 
                      \vrule width.4pt  height3pt }
               \hrule height.4pt width3.8pt
               \hbox {\vrule width.4pt  height3pt  
                      \kern3pt 
                      \vrule width.4pt  height3pt }
               \hrule height.4pt width3.8pt}
                         }
              }
           }
\def\twotwooneoneone{ 
              {\vcenter  {\vbox  
              {\hrule height.4pt
               \hbox {\vrule width.4pt  height3pt  
                      \kern3pt 
                      \vrule width.4pt  height3pt 
                      \kern3pt 
                      \vrule width.4pt  height3pt }
               \hrule height.4pt
               \hbox {\vrule width.4pt  height3pt  
                      \kern3pt 
                      \vrule width.4pt  height3pt 
                      \kern3pt 
                      \vrule width.4pt  height3pt }
               \hrule height.4pt
               \hbox {\vrule width.4pt  height3pt  
                      \kern3pt 
                      \vrule width.4pt  height3pt }
               \hrule height.4pt width3.8pt
               \hbox {\vrule width.4pt  height3pt  
                      \kern3pt 
                      \vrule width.4pt  height3pt }
               \hrule height.4pt width3.8pt
               \hbox {\vrule width.4pt  height3pt  
                      \kern3pt 
                      \vrule width.4pt  height3pt }
               \hrule height.4pt width3.8pt}
                         }
              }
           }
\def\threeoneoneoneone{ 
              {\vcenter  {\vbox  
              {\hrule height.4pt
               \hbox {\vrule width.4pt  height3pt  
                      \kern3pt 
                      \vrule width.4pt  height3pt 
                      \kern3pt 
                      \vrule width.4pt  height3pt 
                      \kern3pt 
                      \vrule width.4pt  height3pt }
               \hrule height.4pt
               \hbox {\vrule width.4pt  height3pt  
                      \kern3pt 
                      \vrule width.4pt  height3pt }
               \hrule height.4pt width3.8pt
               \hbox {\vrule width.4pt  height3pt  
                      \kern3pt 
                      \vrule width.4pt  height3pt }
               \hrule height.4pt width3.8pt
               \hbox {\vrule width.4pt  height3pt  
                      \kern3pt 
                      \vrule width.4pt  height3pt }
               \hrule height.4pt width3.8pt
               \hbox {\vrule width.4pt  height3pt  
                      \kern3pt 
                      \vrule width.4pt  height3pt }
               \hrule height.4pt width3.8pt}
                         }
              }
           }
\def\sixsixtwooneone{ 
              {\vcenter  {\vbox  
              {\hrule height.4pt
               \hbox {\vrule width.4pt  height3pt  
                      \kern3pt 
                      \vrule width.4pt  height3pt 
                      \kern3pt 
                      \vrule width.4pt  height3pt 
                      \kern3pt 
                      \vrule width.4pt  height3pt 
                      \kern3pt 
                      \vrule width.4pt  height3pt 
                      \kern3pt 
                      \vrule width.4pt  height3pt 
                      \kern3pt 
                      \vrule width.4pt  height3pt }
               \hrule height.4pt
               \hbox {\vrule width.4pt  height3pt  
                      \kern3pt 
                      \vrule width.4pt  height3pt 
                      \kern3pt 
                      \vrule width.4pt  height3pt 
                      \kern3pt 
                      \vrule width.4pt  height3pt 
                      \kern3pt 
                      \vrule width.4pt  height3pt 
                      \kern3pt 
                      \vrule width.4pt  height3pt 
                      \kern3pt 
                      \vrule width.4pt  height3pt }
               \hrule height.4pt
               \hbox {\vrule width.4pt  height3pt  
                      \kern3pt 
                      \vrule width.4pt  height3pt 
                      \kern3pt 
                      \vrule width.4pt  height3pt }
               \hrule height.4pt width7.2pt
               \hbox {\vrule width.4pt  height3pt  
                      \kern3pt 
                      \vrule width.4pt  height3pt }
               \hrule height.4pt width3.8pt
               \hbox {\vrule width.4pt  height3pt  
                      \kern3pt 
                      \vrule width.4pt  height3pt }
               \hrule height.4pt width3.8pt}
                         }
              }
           }
\def\twotwotwotwotwotwo{
              {\vcenter  {\vbox  
              {\hrule height.4pt
               \hbox {\vrule width.4pt  height3pt  
                      \kern3pt 
                      \vrule width.4pt  height3pt 
                      \kern3pt 
                      \vrule width.4pt  height3pt }
               \hrule height.4pt
               \hbox {\vrule width.4pt  height3pt  
                      \kern3pt 
                      \vrule width.4pt  height3pt 
                      \kern3pt 
                      \vrule width.4pt  height3pt }
               \hrule height.4pt
               \hbox {\vrule width.4pt  height3pt  
                      \kern3pt 
                      \vrule width.4pt  height3pt 
                      \kern3pt 
                      \vrule width.4pt  height3pt }
               \hrule height.4pt
               \hbox {\vrule width.4pt  height3pt  
                      \kern3pt 
                      \vrule width.4pt  height3pt 
                      \kern3pt 
                      \vrule width.4pt  height3pt }
               \hrule height.4pt
               \hbox {\vrule width.4pt  height3pt  
                      \kern3pt 
                      \vrule width.4pt  height3pt 
                      \kern3pt 
                      \vrule width.4pt  height3pt }
               \hrule height.4pt
               \hbox {\vrule width.4pt  height3pt  
                      \kern3pt 
                      \vrule width.4pt  height3pt 
                      \kern3pt 
                      \vrule width.4pt  height3pt }
               \hrule height.4pt}
                         }
              }
           }

\def\eight{  {\vcenter  {\vbox  
              {\hrule height.4pt
               \hbox {\vrule width.4pt  height3pt  
                      \kern3pt 
                      \vrule width.4pt  height3pt 
                      \kern3pt 
                      \vrule width.4pt  height3pt 
                      \kern3pt 
                      \vrule width.4pt  height3pt 
                      \kern3pt
                      \vrule width.4pt  height3pt 
                      \kern3pt
                      \vrule width.4pt  height3pt 
                      \kern3pt
                      \vrule width.4pt height3pt
                      \kern3pt
                      \vrule width.4pt height3pt
                      \kern3pt
                      \vrule width.4pt height3pt}
               \hrule height.4pt}}}}
\def\sevenone{
	{\vcenter  {\vbox
		{\hrule height.4pt
			\hbox{\vrule width.4pt height3pt
				\kern3pt
				\vrule width.4pt height3pt
				\kern3pt
				\vrule width.4pt height3pt
				\kern3pt
				\vrule width.4pt height3pt
				\kern3pt
				\vrule width.4pt height3pt
				\kern3pt
				\vrule width.4pt height3pt
				\kern3pt
				\vrule width.4pt height3pt
				\kern3pt
				\vrule width.4pt height3pt}
		\hrule height.4pt
			\hbox{\vrule width.4pt height3pt
				\kern3pt
				\vrule width.4pt height3pt}
		\hrule height.4pt width3.8pt}}}}
		
\def\sixtwo{
	{\vcenter {\vbox
		{\hrule height.4pt
			\hbox{\vrule width.4pt  height3pt  
                      	\kern3pt 
                      	\vrule width.4pt  height3pt 
                      	\kern3pt 
                      	\vrule width.4pt  height3pt 
                      	\kern3pt 
                      	\vrule width.4pt  height3pt 
                      	\kern3pt 
                      	\vrule width.4pt  height3pt 
                      	\kern3pt 
                      	\vrule width.4pt  height3pt 
                      	\kern3pt 
                      	\vrule width.4pt  height3pt}
		\hrule height.4pt
			\hbox{\vrule width.4pt  height3pt 
                      	\kern3pt 
                      	\vrule width.4pt  height3pt 
                      	\kern3pt 
                      	\vrule width.4pt  height3pt}
		\hrule height.4pt width7.2pt}}}}
		
\def\fivethree{ 
              {\vcenter  {\vbox  
              {\hrule height.4pt
               \hbox {\vrule width.4pt  height3pt  
                      \kern3pt 
                      \vrule width.4pt  height3pt 
                      \kern3pt 
                      \vrule width.4pt  height3pt 
                      \kern3pt 
                      \vrule width.4pt  height3pt 
                      \kern3pt 
                      \vrule width.4pt  height3pt 
                      \kern3pt
                      \vrule width.4pt height3pt}
               \hrule height.4pt
               \hbox {\vrule width.4pt  height3pt  
                      \kern3pt 
                      \vrule width.4pt  height3pt 
                      \kern3pt
                      \vrule width.4pt height3pt
                      \kern3pt
                      \vrule width.4pt height3pt}
               \hrule height.4pt width10.6pt}}}}

\def\fivetwoone{ 
              {\vcenter  {\vbox  
              {\hrule height.4pt
               \hbox {\vrule width.4pt  height3pt  
                      \kern3pt 
                      \vrule width.4pt  height3pt 
                      \kern3pt 
                      \vrule width.4pt  height3pt 
                      \kern3pt 
                      \vrule width.4pt  height3pt 
                      \kern3pt 
                      \vrule width.4pt  height3pt 
                      \kern3pt 
                      \vrule width.4pt  height3pt }
               \hrule height.4pt
               \hbox {\vrule width.4pt  height3pt  
                      \kern3pt 
                      \vrule width.4pt  height3pt
                      \kern3pt 
                      \vrule width.4pt  height3pt  }
               \hrule height.4pt width7.2pt
               \hbox {\vrule width.4pt  height3pt  
                      \kern3pt 
                      \vrule width.4pt  height3pt }
               \hrule height.4pt width3.8pt}} }}
           
\def\fiveoneoneone{ 
              {\vcenter  {\vbox  
              {\hrule height.4pt
               \hbox {\vrule width.4pt  height3pt  
                      \kern3pt 
                      \vrule width.4pt  height3pt 
                      \kern3pt 
                      \vrule width.4pt  height3pt 
                      \kern3pt 
                      \vrule width.4pt  height3pt 
                      \kern3pt 
                      \vrule width.4pt  height3pt 
                      \kern3pt 
                      \vrule width.4pt  height3pt }
               \hrule height.4pt
               \hbox {\vrule width.4pt  height3pt  
                      \kern3pt 
                      \vrule width.4pt  height3pt }
               \hrule height.4pt width3.8pt
               \hbox {\vrule width.4pt  height3pt  
                      \kern3pt 
                      \vrule width.4pt  height3pt }
               \hrule height.4pt width3.8pt
               \hbox {\vrule width.4pt  height3pt  
                      \kern3pt 
                      \vrule width.4pt  height3pt }
               \hrule height.4pt width3.8pt}}}}

\def\fourfour{ 
              {\vcenter  {\vbox  
              {\hrule height.4pt
               \hbox {\vrule width.4pt  height3pt  
                      \kern3pt 
                      \vrule width.4pt  height3pt 
                      \kern3pt 
                      \vrule width.4pt  height3pt 
                      \kern3pt
                      \vrule width.4pt height3pt
                      \kern3pt
                      \vrule width.4pt height3pt}
               \hrule height.4pt
               \hbox {\vrule width.4pt  height3pt  
                      \kern3pt 
                      \vrule width.4pt  height3pt 
                      \kern3pt
                      \vrule width.4pt  height3pt 
                      \kern3pt
                      \vrule width.4pt height3pt
                      \kern3pt
                      \vrule width.4pt height3pt}
               \hrule height.4pt }}}}
           
\def\fourthreeone{ 
              {\vcenter  {\vbox  
              {\hrule height.4pt
               \hbox {\vrule width.4pt  height3pt  
                      \kern3pt 
                      \vrule width.4pt  height3pt 
                      \kern3pt 
                      \vrule width.4pt  height3pt 
                      \kern3pt 
                      \vrule width.4pt  height3pt 
                      \kern3pt 
                      \vrule width.4pt  height3pt }
               \hrule height.4pt
               \hbox {\vrule width.4pt  height3pt  
                      \kern3pt 
                      \vrule width.4pt  height3pt 
                      \kern3pt 
                      \vrule width.4pt  height3pt 
                      \kern3pt 
                      \vrule width.4pt  height3pt }
               \hrule height.4pt width10.6pt
               \hbox {\vrule width.4pt  height3pt  
                      \kern3pt 
                      \vrule width.4pt  height3pt }
               \hrule height.4pt width3.8pt} }}}
           
\def\fourtwotwo{ 
              {\vcenter  {\vbox  
              {\hrule height.4pt
               \hbox {\vrule width.4pt  height3pt  
                      \kern3pt 
                      \vrule width.4pt  height3pt 
                      \kern3pt 
                      \vrule width.4pt  height3pt 
                      \kern3pt 
                      \vrule width.4pt  height3pt 
                      \kern3pt 
                      \vrule width.4pt  height3pt }
               \hrule height.4pt
               \hbox {\vrule width.4pt  height3pt  
                      \kern3pt 
                      \vrule width.4pt  height3pt 
                      \kern3pt 
                      \vrule width.4pt  height3pt }
               \hrule height.4pt width7.2pt
               \hbox {\vrule width.4pt  height3pt  
                      \kern3pt 
                      \vrule width.4pt  height3pt
                      \kern3pt 
                      \vrule width.4pt  height3pt  }
               \hrule height.4pt width7.2pt}}}}
               
\def\fourtwooneone{ 
              {\vcenter  {\vbox  
              {\hrule height.4pt
               \hbox {\vrule width.4pt  height3pt  
                      \kern3pt 
                      \vrule width.4pt  height3pt 
                      \kern3pt 
                      \vrule width.4pt  height3pt 
                      \kern3pt 
                      \vrule width.4pt  height3pt 
                      \kern3pt 
                      \vrule width.4pt  height3pt }
               \hrule height.4pt
               \hbox {\vrule width.4pt  height3pt  
                      \kern3pt 
                      \vrule width.4pt  height3pt 
                      \kern3pt 
                      \vrule width.4pt  height3pt }
               \hrule height.4pt width7.2pt
               \hbox {\vrule width.4pt  height3pt  
                      \kern3pt 
                      \vrule width.4pt  height3pt }
               \hrule height.4pt width3.8pt
                              \hbox {\vrule width.4pt  height3pt  
                      \kern3pt 
                      \vrule width.4pt  height3pt }
               \hrule height.4pt width3.8pt}}}}
               
\def\fouroneoneoneone{ 
              {\vcenter  {\vbox  
              {\hrule height.4pt
               \hbox {\vrule width.4pt  height3pt  
                      \kern3pt 
                      \vrule width.4pt  height3pt 
                      \kern3pt 
                      \vrule width.4pt  height3pt 
                      \kern3pt 
                      \vrule width.4pt  height3pt 
                      \kern3pt 
                      \vrule width.4pt  height3pt }
               \hrule height.4pt
               \hbox {\vrule width.4pt  height3pt  
                      \kern3pt 
                      \vrule width.4pt  height3pt }
               \hrule height.4pt width3.8pt
               \hbox {\vrule width.4pt  height3pt  
                      \kern3pt 
                      \vrule width.4pt  height3pt }
               \hrule height.4pt width3.8pt
                \hbox {\vrule width.4pt  height3pt  
                      \kern3pt 
                      \vrule width.4pt  height3pt }
               \hrule height.4pt width3.8pt
                \hbox {\vrule width.4pt  height3pt  
                      \kern3pt 
                      \vrule width.4pt  height3pt }
               \hrule height.4pt width3.8pt}}}}
               
\def\threethreetwo{ 
              {\vcenter  {\vbox  
              {\hrule height.4pt
               \hbox {\vrule width.4pt  height3pt  
                      \kern3pt 
                      \vrule width.4pt  height3pt 
                      \kern3pt 
                      \vrule width.4pt  height3pt 
                      \kern3pt 
                      \vrule width.4pt  height3pt }
               \hrule height.4pt
               \hbox {\vrule width.4pt  height3pt  
                      \kern3pt 
                      \vrule width.4pt  height3pt 
                      \kern3pt 
                      \vrule width.4pt  height3pt 
                      \kern3pt 
                      \vrule width.4pt  height3pt }
               \hrule height.4pt width10.6pt
               \hbox {\vrule width.4pt  height3pt  
                      \kern3pt 
                      \vrule width.4pt  height3pt 
                      \kern3pt 
                      \vrule width.4pt  height3pt }
               \hrule height.4pt width7.2pt}}}}
               
\def\threethreeoneone{ 
              {\vcenter  {\vbox  
              {\hrule height.4pt
               \hbox {\vrule width.4pt  height3pt  
                      \kern3pt 
                      \vrule width.4pt  height3pt 
                      \kern3pt 
                      \vrule width.4pt  height3pt 
                      \kern3pt 
                      \vrule width.4pt  height3pt }
               \hrule height.4pt
               \hbox {\vrule width.4pt  height3pt  
                      \kern3pt 
                      \vrule width.4pt  height3pt 
                      \kern3pt 
                      \vrule width.4pt  height3pt 
                      \kern3pt 
                      \vrule width.4pt  height3pt }
               \hrule height.4pt 
               \hbox {\vrule width.4pt  height3pt  
                      \kern3pt 
                      \vrule width.4pt  height3pt }
               \hrule height.4pt width3.8pt
               \hbox {\vrule width.4pt  height3pt  
                      \kern3pt 
                      \vrule width.4pt  height3pt }
               \hrule height.4pt width3.8pt}}}}
               
\def\threetwotwoone{ 
              {\vcenter  {\vbox  
              {\hrule height.4pt
               \hbox {\vrule width.4pt  height3pt  
                      \kern3pt 
                      \vrule width.4pt  height3pt 
                      \kern3pt 
                      \vrule width.4pt  height3pt 
                      \kern3pt 
                      \vrule width.4pt  height3pt }
               \hrule height.4pt
               \hbox {\vrule width.4pt  height3pt  
                      \kern3pt 
                      \vrule width.4pt  height3pt 
                      \kern3pt 
                      \vrule width.4pt  height3pt }
               \hrule height.4pt width7.2pt
               \hbox {\vrule width.4pt  height3pt  
                      \kern3pt 
                      \vrule width.4pt  height3pt 
                      \kern3pt 
                      \vrule width.4pt  height3pt }
               \hrule height.4pt width7.2pt
                \hbox {\vrule width.4pt  height3pt  
                      \kern3pt 
                      \vrule width.4pt  height3pt }
               \hrule height.4pt width3.8pt}}}}

\def\threetwooneoneone{ 
              {\vcenter  {\vbox  
              {\hrule height.4pt
               \hbox {\vrule width.4pt  height3pt  
                      \kern3pt 
                      \vrule width.4pt  height3pt 
                      \kern3pt 
                      \vrule width.4pt  height3pt 
                      \kern3pt 
                      \vrule width.4pt  height3pt }
               \hrule height.4pt
               \hbox {\vrule width.4pt  height3pt  
                      \kern3pt 
                      \vrule width.4pt  height3pt 
                      \kern3pt 
                      \vrule width.4pt  height3pt }
               \hrule height.4pt width7.2pt
               \hbox {\vrule width.4pt  height3pt  
                      \kern3pt 
                      \vrule width.4pt  height3pt }
               \hrule height.4pt width3.8pt
               \hbox {\vrule width.4pt  height3pt  
                      \kern3pt 
                      \vrule width.4pt  height3pt }
               \hrule height.4pt width3.8pt
               \hbox {\vrule width.4pt  height3pt  
                      \kern3pt 
                      \vrule width.4pt  height3pt }
               \hrule height.4pt width3.8pt}}}}
               
\def\threeoneoneoneoneone{ 
              {\vcenter  {\vbox  
              {\hrule height.4pt
               \hbox {\vrule width.4pt  height3pt  
                      \kern3pt 
                      \vrule width.4pt  height3pt 
                      \kern3pt 
                      \vrule width.4pt  height3pt 
                      \kern3pt 
                      \vrule width.4pt  height3pt }
               \hrule height.4pt
               \hbox {\vrule width.4pt  height3pt  
                      \kern3pt 
                      \vrule width.4pt  height3pt }
               \hrule height.4pt width3.8pt
               \hbox {\vrule width.4pt  height3pt  
                      \kern3pt 
                      \vrule width.4pt  height3pt }
               \hrule height.4pt width3.8pt
               \hbox {\vrule width.4pt  height3pt  
                      \kern3pt 
                      \vrule width.4pt  height3pt }
               \hrule height.4pt width3.8pt
               \hbox {\vrule width.4pt  height3pt  
                      \kern3pt 
                      \vrule width.4pt  height3pt }
               \hrule height.4pt width3.8pt
               \hbox {\vrule width.4pt  height3pt  
                      \kern3pt 
                      \vrule width.4pt  height3pt }
               \hrule height.4pt width3.8pt}}}}
               
\def\twotwotwotwo{ 
              {\vcenter  {\vbox  
              {\hrule height.4pt
               \hbox {\vrule width.4pt  height3pt  
                      \kern3pt 
                      \vrule width.4pt  height3pt 
                      \kern3pt 
                      \vrule width.4pt  height3pt }
               \hrule height.4pt
               \hbox {\vrule width.4pt  height3pt  
                      \kern3pt 
                      \vrule width.4pt  height3pt 
                      \kern3pt 
                      \vrule width.4pt  height3pt }
               \hrule height.4pt
               \hbox {\vrule width.4pt  height3pt  
                      \kern3pt 
                      \vrule width.4pt  height3pt 
                      \kern3pt 
                      \vrule width.4pt  height3pt }
               \hrule height.4pt 
               \hbox {\vrule width.4pt  height3pt  
                      \kern3pt 
                      \vrule width.4pt  height3pt 
                      \kern3pt 
                      \vrule width.4pt  height3pt }
               \hrule height.4pt }}}}
               
\def\twotwooneoneoneone{ 
              {\vcenter  {\vbox  
              {\hrule height.4pt
               \hbox {\vrule width.4pt  height3pt  
                      \kern3pt 
                      \vrule width.4pt  height3pt 
                      \kern3pt 
                      \vrule width.4pt  height3pt }
               \hrule height.4pt
               \hbox {\vrule width.4pt  height3pt  
                      \kern3pt 
                      \vrule width.4pt  height3pt 
                      \kern3pt 
                      \vrule width.4pt  height3pt }
               \hrule height.4pt width7.2pt
               \hbox {\vrule width.4pt  height3pt  
                      \kern3pt 
                      \vrule width.4pt  height3pt }
               \hrule height.4pt width3.8pt
               \hbox {\vrule width.4pt  height3pt  
                      \kern3pt 
                      \vrule width.4pt  height3pt }
               \hrule height.4pt width3.8pt
               \hbox {\vrule width.4pt  height3pt  
                      \kern3pt 
                      \vrule width.4pt  height3pt }
               \hrule height.4pt width3.8pt
               \hbox {\vrule width.4pt  height3pt  
                      \kern3pt 
                      \vrule width.4pt  height3pt }
               \hrule height.4pt width3.8pt}}}}

\def\twooneoneoneoneoneone{ 
              {\vcenter  {\vbox  
              {\hrule height.4pt
               \hbox {\vrule width.4pt  height3pt  
                      \kern3pt 
                      \vrule width.4pt  height3pt 
                      \kern3pt 
                      \vrule width.4pt  height3pt }
               \hrule height.4pt
               \hbox {\vrule width.4pt  height3pt  
                      \kern3pt 
                      \vrule width.4pt  height3pt }
               \hrule height.4pt width3.8pt
               \hbox {\vrule width.4pt  height3pt  
                      \kern3pt 
                      \vrule width.4pt  height3pt }
               \hrule height.4pt width3.8pt
               \hbox {\vrule width.4pt  height3pt  
                      \kern3pt 
                      \vrule width.4pt  height3pt }
               \hrule height.4pt width3.8pt
               \hbox {\vrule width.4pt  height3pt  
                      \kern3pt 
                      \vrule width.4pt  height3pt }
               \hrule height.4pt width3.8pt
               \hbox {\vrule width.4pt  height3pt  
                      \kern3pt 
                      \vrule width.4pt  height3pt }
               \hrule height.4pt width3.8pt
               \hbox {\vrule width.4pt  height3pt  
                      \kern3pt 
                      \vrule width.4pt  height3pt }
               \hrule height.4pt width3.8pt}}}}
               
\def\oneupeight{  {\vcenter  {\vbox  
              {\hrule height.4pt
               \hbox {\vrule width.4pt  height3pt  
                      \kern3pt 
                      \vrule width.4pt  height3pt }
               \hrule height.4pt
               \hbox {\vrule width.4pt  height3pt  
                      \kern3pt 
                      \vrule width.4pt  height3pt }
               \hrule height.4pt
               \hbox {\vrule width.4pt  height3pt  
                      \kern3pt 
                      \vrule width.4pt  height3pt }
               \hrule height.4pt
               \hbox {\vrule width.4pt  height3pt  
                      \kern3pt 
                      \vrule width.4pt  height3pt }
               \hrule height.4pt
               \hbox {\vrule width.4pt  height3pt  
                      \kern3pt 
                      \vrule width.4pt  height3pt }
               \hrule height.4pt
               \hbox {\vrule width.4pt  height3pt  
                      \kern3pt 
                      \vrule width.4pt  height3pt }
               \hrule height.4pt
               \hbox {\vrule width.4pt  height3pt  
                      \kern3pt 
                      \vrule width.4pt  height3pt }
               \hrule height.4pt
               \hbox {\vrule width.4pt  height3pt  
                      \kern3pt 
                      \vrule width.4pt  height3pt }
               \hrule height.4pt}
                         }
                   }
           }

\begin{document}

\begin{flushright}
BOW-PH-155\\
\end{flushright}
\vspace{3mm}

\begin{center}
{\Large\bf\sf  
Symmetric-group decomposition of \\
SU($N$) group-theory constraints on \\
four-, five-, and six-point color-ordered amplitudes  \\ [2mm]
at all loop orders
}

\vskip 1.5cm 

{\sc
Alexander C. Edison 
and 
Stephen G. Naculich\footnote{ 
Research supported in part by the National Science 
Foundation under Grant No.~PHY10-67961.}
}

\vskip 0.5cm 
{\it 
Department of Physics\\
Bowdoin College\\
Brunswick, ME 04011, USA
}

\vspace{5mm}
{\tt 
aedison@bowdoin.edu\\
naculich@bowdoin.edu
}
\end{center}

\vskip 1.5cm

\begin{abstract}

Color-ordered amplitudes for the scattering of 
$n$ particles in the adjoint representation 
of SU($N$) gauge theory satisfy constraints
that arise from group theory alone.   
These constraints break into subsets associated with
irreducible representations of the symmetric group $S_n$,
which allows them to be presented in a compact and natural way.
Using an iterative approach, we derive the constraints 
for six-point amplitudes at all loop orders,
extending earlier results for $n=4$ and $n=5$.
We then decompose the four-, five-, and six-point group-theory 
constraints into their irreducible $S_n$  subspaces.
We comment briefly on higher-point two-loop amplitudes.

\end{abstract}

\vfil\break

\section{Introduction}
\setcounter{equation}{0}

Scattering amplitudes in gauge theory 
may be given a gauge-invariant decomposition 
in terms of color-ordered (or partial) 
amplitudes \cite{Mangano:1990by,Bern:1990ux}.
Color-ordered amplitudes are not independent but satisfy a number 
of constraints, some of which follow from the recently-discovered 
color-kinematic 
duality \cite{Bern:2008qj,Bern:2010ue,Carrasco:2011hw}
and were proven in 
refs.~\cite{BjerrumBohr:2009rd,Stieberger:2009hq,Feng:2010my,Chen:2011jxa}.
Other constraints among the color-ordered amplitudes, however, 
follow directly from group theory, 
and have been known for over two decades at 
tree-level \cite{Kleiss:1988ne} 
and at one loop \cite{Bern:1994zx}.
Four-point color-ordered amplitudes 
were also known to obey group-theory relations at two loops \cite{Bern:2002tk}
and these were recently generalized to all loop orders \cite{Naculich:2011ep}.
All-loop-order group-theory constraints were also recently derived 
for five-point amplitudes \cite{Edison:2011ta}. 
Other recent work on loop-order relations among color-ordered amplitudes
includes 
refs.~\cite{BjerrumBohr:2011xe,Feng:2011fja,Boels:2011tp,Boels:2011mn,Du:2011se,Du:2012mt}.

In this paper, we derive the SU($N$) group-theory constraints 
for six-point color-ordered amplitudes at all loop orders,
following the iterative approach of refs.~\cite{Naculich:2011ep,Edison:2011ta}
and generalizing the known relations at tree level \cite{Kleiss:1988ne} and 
at one loop \cite{Bern:1994zx}.
The other major goal of this paper is to 
describe how group-theory constraints on $n$-point color-ordered amplitudes
naturally fall into sets associated with irreducible representations 
of the symmetric group $S_n$.
We decompose the constraints 
for $n=4$, 5, and 6 into their irreducible $S_n$  subspaces,
thus presenting them in a compact way.

The derivation of group-theory constraints can be cast
as a straightforward problem in linear algebra\footnote{
A similar approach was taken to the BCJ relations \cite{Bern:2008qj}
in ref.~\cite{Vaman:2010ez}.}
that emerges from two alternative ways of expressing 
the color structure of a gauge theory amplitude.
One way is to decompose the amplitude into a color
basis \cite{DelDuca:1999ha,DelDuca:1999rs}
\be
\cA= \sum_i  a_i c_i
\label{colorbasis}
\ee
where $a_i$ carries the momentum- and polarization-dependence 
of the amplitude,
and the color factors $c_i$ are obtained by sewing together
the gauge theory factors from all the vertices of
the contributing Feynman diagrams.
In a theory that contains only fields 
in the adjoint representation of SU($N$), 
such as pure or supersymmetric Yang-Mills theory, 
each cubic vertex contributes a factor of the 
SU($N$) structure constants $f^{abc}$,
and each quartic vertex contributes factors of 
$f^{abe} f^{cde}$,  $f^{ace} f^{bde}$, and $f^{ade} f^{bce}$,
which are equivalent from a purely color point-of-view
to a pair of cubic vertices sewn along one leg.
Hence a complete set
of color factors $\{ c_i \}$ 
can be constructed from diagrams
with cubic vertices only.
An independent basis of color factors\footnote{
The color factors constructed from the set of all cubic diagrams 
are generally not independent but are related by Jacobi relations. 
Such an overcomplete set is usually required 
to make color-kinematic duality manifest\cite{Bern:2008qj,Carrasco:2011mn}. 
}
for tree-level and one-loop $n$-point amplitudes 
was described in refs.~\cite{DelDuca:1999ha,DelDuca:1999rs}.

The alternative trace decomposition \cite{Mangano:1990by,Bern:1990ux}
\be
\cA= \sum_\la  \Ala   t_\la 
\label{tracebasis}
\ee
expresses the amplitude in terms of 
gauge-invariant color-ordered amplitudes $\Ala$
with respect to a basis $\{  t_\la \}$ of single and (at loop level) multiple 
traces of gauge group generators
$T^a$ in the defining  representation of SU($N$).
We will characterize this basis more explicitly in sec.~\ref{sec-trace}.
The dimension of the trace basis, however, is larger than
that of the independent color basis\footnote{
For example, the number of independent color factors 
for tree-level $n$-point amplitudes is $(n-2)!$ 
while the dimension of the trace basis is $\half (n-1)!$.
},
so there is redundancy among the color-ordered amplitudes.

The color (\ref{colorbasis}) and trace (\ref{tracebasis})
decompositions can be related by using\footnote{
The generators are normalized using $\Tr (T^a T^b) = \delta^{ab}$.
}
\be
\f^{abc} =
i \sqrt2 f^{abc} = \Tr( [T^a, T^b] T^c ) \,,
\label{structure}
\ee
together with the SU($N$) relations
given later in \eqn{sunrelations}, 
to write each color factor $c_i$  as a linear combination of trace factors
\be
c_i =  \sum_\la \Mila t_\la \,.
\label{trans}
\ee
Since the dimension of the trace space 
is larger than the number of independent color factors, 
the linear combinations given by \eqn{trans} 
span a proper subspace 
(which we will refer to as the {\it color space})  
of the trace space. 
Consequently, the transformation matrix $\Mila$ possesses 
a set of independent null eigenvectors 
\be
\sum_\la  \Mila  r_{\la \aa} = 0 \,, \qquad  \aa = 1, \cdots, \ncon 
\label{null}
\ee
whose number $\ncon$ is the difference between the dimensions of the trace
space and the color space.  The null vectors 
$r_\aa = \sum_\la r_{\la \aa} t_\la $ 
are orthogonal to the color factors $c_i$ with respect to the inner
product 
\be 
(t_\la, t_{\la'} ) = \delta_{\la\la'} 
\label{innerproduct}
\ee 
and hence span the orthogonal complement of the color space, which
we will refer to as the {\it null space}.
One combines \eqn{trans} with \eqns{colorbasis}{tracebasis}
to express the color-ordered amplitudes as
\be
\Ala = \sum_i  a_i \Mila \,.
\label{colorordered}
\ee
The existence of the null eigenvectors (\ref{null}) 
implies a set of constraints
\be
\sum_\la  \Ala r_{\la \aa} = 0 \,, \qquad \aa = 1, \cdots, \ncon
\label{rightrelations}
\ee
which we refer to as group-theory relations.
Hence, specifying the null space is equivalent to specifying the complete
set of group-theory relations satisfied by the color-ordered amplitudes.

For tree-level $n$-point amplitudes, 
the relations (\ref{rightrelations}), which number $\half (n-3) (n-2)!$,
are the Kleiss-Kuijf \cite{Kleiss:1988ne,DelDuca:1999rs} relations.
For one-loop $n$-point amplitudes,
the group-theory relations were given in refs.~\cite{Bern:1994zx,DelDuca:1999rs}.
At two loops and above, the full set of 
constraints on $n$-point color-ordered amplitudes has 
yet to be identified\footnote{
See refs.~\cite{Bern:2002tk,Feng:2011fja} for partial results at two loops.},
although all-loop group-theory relations for $n=4$ and $n=5$ 
have been derived using an iterative approach in 
refs.~\cite{Naculich:2011ep,Edison:2011ta}.

For {\it four-point amplitudes},
there are 4 group-theory relations at each loop order $L \ge 2$   
(with 1 at tree level, and 3 at one loop). 
The form of the constraints begins to repeat after three loops,
so that the even-loop constraints for $L\ge 4$ 
are essentially equivalent to the two-loop constraints,
and the odd-loop constraints for $L\ge 5$ are equivalent to the 
three-loop constraints \cite{Naculich:2011ep}.

For {\it five-point amplitudes}, 
there are 10 group-theory relations for odd $L$,
and  12 for even $L \ge 2$ (with 6 at tree level). 
Again the form of the constraints repeats after two loops,
so that 
odd-loop constraints for  $L \ge 3$ 
are equivalent to the one-loop constraints,
and even-loop constraints for  $L \ge 4$ 
are equivalent to the two-loop constraints  \cite{Edison:2011ta}.

In this paper we will derive the group-theory relations for 
{\it six-point amplitudes} at all loop orders using the same iterative approach.
We will find that
there are 76 group-theory relations for $L \ge 3$, 
with 36 at tree level, 65 at one loop, and 80 at two loops. 
Once again the form of the constraints presents a repeating pattern,
which only begins, however, after five loops,
so that even-loop constraints for $L \ge 6$ are equivalent 
to the four-loop constraints, 
and odd-loop constraints for $L \ge 7$ are equivalent 
to the five-loop constraints. 

A subset of the group-theory relations obeyed by color-ordered amplitudes
can be derived as U(1) decoupling 
relations \cite{Berends:1987cv,Mangano:1987xk,Berends:1987me,Bern:1990ux}.
To obtain these, one enlarges the gauge group from SU($N$) to U($N$).
The trace decomposition is expanded to include terms with factors of $\Tr(T^a)$ 
(which automatically vanish in the SU($N$) trace decomposition).
Next, one observes that an $n$-point amplitude containing any number of external
U(1) gauge bosons vanishes because the associated structure constants are zero. 
By setting the corresponding U($N$) generators in the full U($N$) trace decomposition 
equal to the unit matrix, one obtains a series of equations 
that relate the remaining SU($N$) amplitudes.
These are the U(1) decoupling relations.
Throughout the paper, we will indicate which of the group-theory relations
can be obtained from U(1) decoupling relations
and which cannot. 

The group-theory relations at four and five points were
presented in refs.~\cite{Naculich:2011ep,Edison:2011ta} 
by specifying a complete set of $L$-loop null vectors 
$\{ r^\Ell_\aa \}$ in terms of their components $r^\Ell_{\la \aa}$ 
with respect to an explicit $L$-loop trace basis $\{ t^\Ell_\la \}$.
While this approach has the advantage of explicitness, 
it quickly becomes unwieldy 
as the dimensions of both the null space and the trace space
grow with $n$.
For example, listing the 200 coefficients of the 80 null vectors
for two-loop six-point amplitudes would hardly be enlightening.
Clearly, a more streamlined approach is called for.

Such a compact approach for presenting the null vectors of $n$-point
amplitudes is provided by the representation 
theory of the symmetric group $S_n$.
Since a permutation of the arguments of an element
of the trace basis $t_\la$ for $n$-point amplitudes 
yields another element of the trace basis,
the trace space constitutes a (reducible) representation of 
the symmetric group $S_n$.    
Similarly, a permutation of the external legs of a diagram
corresponding to a color factor $c_i$ yields a diagram 
corresponding to another color factor, 
hence the color space constitutes an invariant subspace 
(with respect to permutations) of the trace space.
The orthogonal complement of the color space,
namely the null space, is therefore also an invariant subspace,
since the inner product (\ref{innerproduct})
is invariant under permutations.
Thus the null space forms a (reducible) representation of $S_n$,
which can be decomposed into a direct sum of irreducible representations.
Indeed, the null vectors for five-point amplitudes found in ref.~\cite{Edison:2011ta}
naturally broke into sets of 6 and 4, 
and correspond to irreducible representations of $S_5$.
Thus, a compact and efficient way to characterize the null space
is to describe its constituent irreducible representations.

In this paper, we will present the null spaces of four-, five-, and
six-point amplitudes at
all loop orders in terms of their irreducible subspaces with respect to $S_n$.
We will find, for example, that the 80 vectors 
spanning the null space for two-loop six-point amplitudes
can be classified into 13 irreducible representations whose
dimensions vary from 1 to 16.
In addition to compactness, this method 
has the advantage that the results are
presented in a form independent of the specific choice of trace basis.

We also briefly examine whether the 
tree-level Kleiss-Kuijf relations also apply to
subleading-color single-trace $n$-point amplitudes at loop level.
We confirm the results of ref.~\cite{Feng:2011fja} that at two loops,
the KK relations hold for $n \le 7$, but fail for eight-point amplitudes.
We check that they fail for nine-point amplitudes as well.

We begin in section \ref{sec-trace} 
with an explicit look at the trace basis for $n$-point 
amplitudes through two loops. 
In section \ref{sec-symmetric}, we review some representation theory 
of the symmetric group, showing how Young tableaux can be used to construct
projection operators onto irreducible representations
for regular and induced representations.
In sections 
\ref{sec-four}, \ref{sec-five}, and  \ref{sec-six},
we describe the decomposition of the 
null spaces of four-, five-, and 
six-point amplitudes into irreducible representations of $S_n$.
Section \ref{sec-KK} discusses Kleiss-Kuijf relations at loop level, and 
section \ref{sec-concl} contains our conclusions. 
Various additional details can be found in two appendices.

\section{Trace bases for gauge-theory amplitudes}
\setcounter{equation}{0}
\label{sec-trace}

In this section, 
we spell out the explicit form of the trace basis 
for an $n$-point amplitude through two loops
as a prelude to examining
its decomposition into irreducible representations of $S_n$
later in the paper.

First we consider tree-level $n$-point amplitudes $\cA^\Zero_n$.
One can use \eqn{structure} together with $\f^{abc} T^a = [T^b, T^c] $
to decompose the color factor $c_i$ built from any tree-level color diagram
into a linear combination of single-trace terms,
so that the amplitude may be expressed as \cite{Mangano:1990by}
\be
\cA^\Zero_n =
\sum_{\sigma \in S_n/\IZ_n}
A( \sig(1), \cdots, \sig(n) )
\Tr( \sig(1) \cdots \sig(n) )
\label{treelevelamp}
\ee
where $\IZ_n$ denotes the subgroup of cyclic permutations.
For easier readability we write 
$\Tr(12\cdots n)$
for 
$\Tr\left(  T^{a_1} T^{a_2} \cdots T^{a_n} \right)  $.
In addition to being invariant under cyclic
permutations of their arguments (due to the cyclicity of the trace),
color-ordered tree amplitudes 
are invariant (up to sign) under reversal of the arguments
\be 
A( \sig(n), \cdots,\sig(2), \sig(1) )=
(-1)^n  A(  \sig(1),\sig(2), \cdots, \sig(n) )
\ee
as can be seen by using the antisymmetry of $\f^{abc}$. 
Hence, the tree-level $n$-point amplitude can be written
\be
\cA_n^\Zero  = \sum_{\la=1}^{\nsing}  \Ala^\Zero   T_\la
\ee
in terms of a basis $\{ T_\la \}$ 
of dimension $\nsing = \half (n-1)!$,
consisting of sums/differences of the form 
\be
\Tr  (  12 \cdots n ) + (-1)^n \Tr  (  n \cdots 21 ) 
\ee
and permutations thereof.

At one loop and above, the amplitude also contains 
double-trace terms 
\be
A (\sig(1), \cdots, \sig(p) \,;\, \sig(p+1), \cdots \sig(n) )
\ \Tr (\sig(1) \cdots \sig(p) ) \ \Tr (\sig(p+1) \cdots \sig(n) ),
\qquad \sig \in S_n
\ee
with  $2 \le p \le \lfloor n/2 \rfloor$.
The color-ordered amplitudes again satisfy a reflection property
\be 
A ( \sig(p), \cdots, \sig(1) \, ; \, \sig(n), \cdots, \sig(p+1) )=
(-1)^n  A (  \sig(1), \cdots, \sig(p) \, ; \,  \sig(p+1),  \cdots ,\sig(n) )
\label{reflection}
\ee
allowing the amplitude to be written in terms of a basis containing terms of the form 
\be
\Tr\left(  1 \cdots p \right) 
\Tr\left( (p+1) \cdots n \right) +  (-1)^n 
\Tr\left(  p \cdots 1    \right) 
\Tr\left(  n \cdots (p+1)  \right) 
\label{doubletraceterm}
\ee
and permutations thereof,
which we label as $ \{ T_{\la} \}$ 
with $\la = \nsing + 1, \cdots, \nsing+ \ndoub$.
The number of independent terms of the form (\ref{doubletraceterm}) 
is $\ndoub = \fourth \sum_{p=2}^{n-2}   n!/p(n-p)$,
except for $n=4$, in which case there are 3 such terms.
The full one-loop $n$-point amplitude can then be expressed 
as \cite{Bern:1990ux}
\be
\cA^\One_n  = \sum_{\la=1}^{\nsing+\ndoub}  \Ala^\One   t^\One_\la 
,\qquad\
t^\One_\la = \begin{cases}
N T_\la, & \la = 1, \cdots, \nsing \\
T_{\la} , & \la = \nsing + 1, \cdots, \nsing+\ndoub \,.
\end{cases}
\ee

Beginning at two loops (and $n \ge 6$),
the trace basis also requires
triple-trace terms of the form
\be
\Tr\left(  1 \cdots p \right)
\Tr\left( (p+1) \cdots q \right) 
\Tr\left( (q+1) \cdots n \right) 
+   (-1)^n 
\Tr\left(  p \cdots 1 \right)
\Tr\left( q \cdots (p+1) \right) 
\Tr\left( n  \cdots (q+1) \right) 
\ee
and permutations thereof, 
which we label as $ \{ T_{\la} \}$ 
with $\la = \nsing + \ndoub + 1, \cdots, \nsing+ \ndoub+ \ntrip$.
The full two-loop $n$-point amplitude can be written 
\be
\cA^\Two_n  = \sum_{\la=1}^{2 \nsing+\ndoub+\ntrip}  \Ala^\Two   t^\Two_\la 
,\qquad\
t^\Two_\la = \begin{cases}
N^2 T_\la, & \la = 1, \cdots, \nsing \\
N T_{\la} , & \la = \nsing + 1, \cdots, \nsing+\ndoub \\
 T_{\la} , & \la = \nsing + \ndoub + 1, \cdots, \nsing+\ndoub + \ntrip \\
 T_{\la-\nsing-\ndoub-\ntrip} , & \la = \nsing + \ndoub + \ntrip + 1, 
\cdots, 2 \nsing+\ndoub + \ntrip 
\end{cases}
\ee
which includes subleading-color single-trace amplitudes,
as well as leading-color single-trace amplitudes
and double- and triple-trace amplitudes.

One can continue this procedure at higher loops,
where the $L$-loop trace basis will include
(for general $n$) up to $(L+1)$-trace terms,
as  well as additional subleading-color terms. 
In secs.~\ref{sec-four}-\ref{sec-six}, 
we will characterize the general $L$-loop basis
for $n=4$, 5, and 6,
which only require up to triple-trace terms.

We now turn to the decomposition of the single- and multiple-trace
bases into irreducible representation of $S_n$.

\section{Projection operators for representations of $S_n$} 
\setcounter{equation}{0}
\label{sec-symmetric}

In this section, we review some standard results 
from the representation theory of the symmetric group $S_n$,
which can be found in many group theory textbooks.  
We found ref.~\cite{Schensted} particularly useful.
First we will describe a method to project the reducible regular representation
onto its irreducible subspaces.
Then we will turn our attention to the induced representations
spanned by the trace bases described in sec.~\ref{sec-trace}.

\subsection{Regular representation}
\label{sec-regular}

We begin by recalling that irreducible representations of 
$S_n$ are labeled by $\tau \in Y_n$,
where $Y_n$ denotes the set of Young tableaux with $n$ boxes.
The dimension of the representation labeled by $\tau$ is given
by $ d_\tau =  n! / H$,
where $H$ is the product of all the hook lengths
associated with $\tau$.

The $n!$-dimensional regular representation of $S_n$ is reducible,
and contains all of the irreducible representations $\tau$  of $S_n$,
each with a multiplicity equal to $d_\tau$,
which implies that $\sum_{\tau \in Y_n} d^2_\tau = n!$.
For example, the regular representation of 
$S_4$ reduces into
\be
{R^\reg_{S_4} \atop {\rm dim}=24}   
\qquad
{= \atop =} 
\qquad
{\four \atop 1}  \plus 
{3~ \threeone \atop 3 \cdot 3} \plus 
{2~\twotwo \atop 2 \cdot 2}  \plus 
{3~\twooneone \atop 3 \cdot 3}  \plus
{\oneoneoneone \atop 1}  \,.
\ee
A basis $\{ f_\sigma \}$ for the regular representation of $S_n$
can be constructed from an arbitrary function 
of $n$ variables
$f(x_1,x_2,\cdots,x_n)$
by permuting the arguments 
\be
f_\sigma \equiv f(x_{\sigma(1)},x_{\sigma(2)}, \cdots,x_{\sigma(n)}),
\qquad\qquad
\sigma \in S_n    \,.
\ee 
Then a permutation $\rho \in S_n$
acts on $f_\sigma$ 
by\footnote{
Our convention is that $\rho\cdot\sigma$ denotes the 
permutation obtained by acting first with $\sigma$ and then with $\rho$.
Hence if $\sigma=(12)$ and $\rho=(23)$ then $\rho\cdot\sigma = (132)$,
where we use the cycle notation for permutations.
Note that the permutations act on the labels rather than the 
positions of the arguments, so that 
$(123) \cdot f(x_4, x_3, x_2, x_1) = f(x_4, x_1, x_3, x_2)$
and not $f(x_2, x_4, x_3, x_1)$. 
}
\be
\rho \cdot f_\sigma \equiv f_{\rho\cdot \sigma} 
= \sum_{\sigma' \in S_n}  f_{\sigma'} D^\reg_{\sigma'\sigma} (\rho)
\ee
where $D^\reg_{\sigma'\sigma} (\rho)$
are the matrices of the regular representation.
We can put 
$D^\reg_{\sigma'\sigma} (\rho)$
into block diagonal form 
by choosing a different basis $\{ f^\tau_{ij} \}$ defined by
\be
f^\tau_{ij} = 
e^\tau_{ij} \cdot f(x_1,x_2,\cdots,x_n), \qquad\qquad
\tau \in Y_n, \quad i,j = 1, \cdots, d_\tau
\ee
where $\{ e^\tau_{ij} \}$ are a set of $n!$ elements
of the group algebra (i.e., linear combinations of permutations)
that satisfy (for each $\tau$) the simple matrix algebra
\be
e_{kl}^\tau e_{ij}^{\tau'} = \delta^{\tau\tau'}  \delta_{li} e_{kj}^\tau \,.
\label{simatalg}
\ee
A permutation $\rho$ acts on this basis as\footnote{
\Eqn{irredrep} follows from \eqn{simatalg} and the fact that any 
$\rho$ is a linear combination of the $e_{kl}^\tau $.}
\be
\rho \cdot f_{ij}^\tau  = \sum_{k=1}^{d_\tau} 
f_{kj}^\tau D_{ki}^{\tau} (\rho)
\label{irredrep}
\ee
showing that each subset of functions
\be
\{       f_{ij}^\tau ~  | ~  i = 1, \cdots, d_\tau \}
\ee
spans an irreducible subspace of the regular representation.
Hence,  $e_{ij}^\tau$ act as projection operators from the regular 
representation onto irreducible representations of $S_n$.
The index $j=1, \cdots,  d_\tau$ labels distinct copies of the 
representation $\tau$, 
each of which transforms according to the same set of
matrices $D_{ki}^{\tau} (\rho)$.

Explicit expressions for $e^\tau_{ij}$ may be constructed 
from Young tableaux as follows \cite{Littlewood,Schensted}.
For any representation $\tau$, 
define the set of standard tableaux 
$\tau_i$,   $i=1, \cdots, d_\tau$,
using the ordering given on p. 139 of ref.~\cite{Schensted}.
Then for each standard tableau $\tau_i$, define the row symmetrizer $P_i$
as the sum of permutations that exchange only symbols within the same row,
and the column antisymmetrizer $Q_i$
as the (signed) sum of permutations that exchange only symbols 
within the same column, where the sign is given by the parity of the
permutation.
Finally, define $s_{ij}$ as the permutation of symbols that converts
tableau $\tau_j$ into $\tau_i$. 
For example, the standard tableaux 
for the $\twotwo$ representation of $S_4$ are given by
\be
\tau_1 = {\def\lr#1{\multicolumn{1}{|@{\hspace{.6ex}}c@{\hspace{.6ex}}|}{\raisebox{-.3ex}{$#1$}}}
\raisebox{-.6ex}{$\begin{array}[b]{cc}
\cline{1-1}\cline{2-2}
\lr{1}&\lr{2}\\
\cline{1-1}\cline{2-2}
\lr{3}&\lr{4}\\
\cline{1-1}\cline{2-2}
\end{array}$}
}
\qquad\qquad
\tau_2 = 
{\def\lr#1{\multicolumn{1}{|@{\hspace{.6ex}}c@{\hspace{.6ex}}|}{\raisebox{-.3ex}{$#1$}}}
\raisebox{-.6ex}{$\begin{array}[b]{cc}
\cline{1-1}\cline{2-2}
\lr{1}&\lr{3}\\
\cline{1-1}\cline{2-2}
\lr{2}&\lr{4}\\
\cline{1-1}\cline{2-2}
\end{array}$}
}
\label{stantabtwotwo}
\ee
so that 
\ba
P_1 = 1 + (12) + (34) + (12)(34), \qquad
Q_1 = 1 - (13) - (24) + (13)(24), \qquad
s_{12} =  (23)\,, \nn\\
P_2 = 1 + (13) + (24) + (13)(24), \qquad
Q_2 = 1 - (12) - (34) + (12)(34), \qquad
s_{21} =  (23)  \,.
\ea
With these definitions, one can construct
projection operators
\be
e_{ij}^\tau  = \left( d_\tau \over n!\right)  Q_i s_{ij} P_j  M_j, \qquad\qquad
M_j = 1 - \sum_{k=1}^{j-1} e_{kk}^\tau
\label{young}
\ee
that obey the simple matrix algebra (\ref{simatalg}).
The diagonal elements $e^\tau_{ii}$ are 
mutually annulling idempotents
\be
e_{kk}^\tau e_{ii}^{\tau'} = \delta^{\tau\tau'}  \delta_{ki} e_{ii}^\tau
\ee
known as Young operators.

In the next subsection, we show that $e^\tau_{ij}$ 
can also be used to project single- and multiple-trace spaces
onto irreducible representations  of $S_n$.

\subsection{Induced representations} 
\label{sec-induced}

As discussed in sec.~\ref{sec-trace}, gauge theory amplitudes can be expressed
in terms of a basis of single and multiple traces of SU($N$) generators.
Since these sets of traces are closed under permutations of their arguments,
they form representations of the symmetric group.    
These single- and multiple-trace representations are induced 
by one-dimensional representations of various subgroups of the symmetric group.
They are in general reducible, and in this section we describe
how to determine their decomposition into irreducible representations,
and how to use the projection operators introduced in sec.~\ref{sec-regular}
to find their irreducible subspaces.

First consider an element of the single-trace basis
introduced in sec.~\ref{sec-trace}
\be
\Tr  (  12 \cdots n ) + (-1)^n \Tr  (  n \cdots 21 )  \,.
\label{tracebasiselement}
\ee
Permutations act on \eqn{tracebasiselement} 
in the same way as on $f(x_1, \cdots, x_n)$, namely
\be
\rho \cdot 
\Big[ \Tr\left(12 \cdots n \right) +  (-1)^n 
\Tr\left( n \cdots 21 \right) \Big]
 ~=~
\Tr\left(\rho(1)\rho(2) \cdots \rho(n) \right) +  (-1)^n 
\Tr\left( \rho(n) \cdots \rho(2)\rho(1) \right)
\ee
thus generating a (reducible) representation of $S_n$,
which we will refer to as the single-trace representation $R^\ind_n$.
\Eqn{tracebasiselement} is invariant\footnote{up to a sign when $n$ is odd}
under the dihedral subgroup $D_n \subset S_n$ of order $2n$
(generated by cyclic permutations and reversal of indices),
so the dimension of $R^\ind_n$ is $\half (n-1)!$.
The superscript $\ind$ denotes that $R^\ind_n$ is the
representation induced by a one-dimensional representation 
of the dihedral group.
When $n$ is even, $R^\ind_n$ is induced by 
the trivial representation of the dihedral subgroup;
when $n$ is odd, it is induced by the representation of 
the dihedral subgroup that assigns $+1$ to cyclic permutations 
and $-1$ to reversals.    

The single-trace representation $R^\ind_n$ is a subset of the 
regular representation, and contains $m_\tau \le d_\tau$
copies of the irreducible representation $\tau$ of $S_n$,
where the multiplicities $m_\tau$ 
may be determined by character analysis (see appendix \ref{app-char}).
As in sec.~\ref{sec-regular}, 
we can explicitly construct the irreducible subspaces 
of $R^\ind_n$ using the projection operators (\ref{young}).
For each $\tau \in Y_n$, 
we construct a basis of the associated irreducible representation(s) via
\be
\repij{n}{\tau}{j} 
\equiv
e_{ij}^\tau \left[  \Tr(1 \cdots n) + (-1)^n \Tr(n \cdots 1) \right], 
\qquad\qquad i = 1, \cdots, d_\tau \,.
\ee
For each $\tau$, there are several sets of basis elements,
labeled by $j=1, \cdots, d_\tau$,
but if $\tau$ appears $m_\tau$ times in $R^\ind_n$, 
then only $m_\tau$ of these sets will be linearly independent,
and we will use only the first $m_\tau$ (independent) values of $j$.
(If $m_\tau=0$, then all of the associated $\repij{n}{\tau}{j}$  vanish.)
If $m_\tau=1$, 
then we will use the lowest value of $j$ 
for which $ v_i \left( n |  \tau , j \right) $ does not vanish,
generally $j=1$, 
and will omit $j$ from the argument, 
i.e.  $ v_i \left( n |  \tau \right) $.

Let us illustrate this with an example. 
Consider the four-point single-trace
representation $R^\ind_4$ spanned by 
\be
\{ \Tr(1234) + \Tr(1432) , \qquad
   \Tr(1243) + \Tr(1342), \qquad
   \Tr(1324) + \Tr(1423) \} \,.
\ee
{}Using character analysis, as described in appendix \ref{app-char}, we find that
this reduces into 
\be
{R^\ind_4 \atop 3} \equals {\four \atop 1 }  \plus {\twotwo \,\,.\atop 2}
\ee
Since $ e^\four = { 1 \over 24 } \sum_{\sigma \in S_4} \sigma $,
the one-dimensional subspace 
corresponding to $\four$ consists of 
the completely symmetric sum of single-trace terms
\ba
\rep{4}{\four}
&=&
e^\four ~\big[ \Tr(1234)  + \Tr(4321)\big]
\label{fourvertfour} 
\\
&=& 
\third  \big[ \Tr(1234)  + \Tr(1432) 
     +  \Tr(1243)  + \Tr(1342)
      + \Tr(1324)  + \Tr(1432)
	\big]
\nn
\ea
while the two-dimensional subspace corresponding to $\twotwo$ is
spanned by 
\be
\repi{4}{\twotwo} =
e_{i1} ^\twotwo ~\big[ \Tr(1234)  + \Tr(4321)\big]
= 
\begin{cases}
\third  \left[ \Tr(1243) + \Tr(1342) - \Tr(1324) - \Tr(1423) \right], & i=1\,,\\
\\
\third  \left[ \Tr(1243) + \Tr(1342) - \Tr(1234) - \Tr(1432) \right], & i=2\,.
\end{cases}
\label{fourverttwotwo} \ee As explained above, since $\twotwo$ appears
only once in $R^\ind_4$,  
$\repij{4}{\twotwo}{2}$ is
proportional to $\repi{4}{\twotwo} \equiv \repij{4}{\twotwo}{1}$.

Next we consider double-trace representations. 
An element of the form
\be
\Tr\left(  1 \cdots p \right) 
\Tr\left( (p+1) \cdots n \right) +  (-1)^n 
\Tr\left(  p \cdots 1    \right) 
\Tr\left(  n \cdots (p+1)  \right)
\label{doubletraceagain}
\ee
is invariant (up to sign when $n$ is odd) 
under the subgroup of $S_n$ generated by cyclic permutations 
of the arguments of each trace and by simultaneous reversal 
of the arguments within each trace  
(and, if $p=n/2$, by an exchange of the two traces).  
We will denote by $R^\ind_{n; p-n}$
the $S_n$ representation induced 
by the representation of this subgroup 
that leaves \eqn{doubletraceagain} invariant.

The multiplicities of the irreducible representations $\tau$
contained in $R^\ind_{n; p-n}$ may again be obtained by character analysis,
and the irreducible subspace(s) corresponding to $\tau$  are spanned by 
\be
\repij{p;n-p}{\tau}{j}  
\equiv
e_{ij}^\tau 
~\big[ 
\Tr\left( 1\cdots p\right) \Tr\left( (p+1)\cdots n \right) +  (-1)^n 
\Tr\left(  p\cdots 1 \right) \Tr\left(  n\cdots (p+1) \right)
\big] \,.
\ee
For example, the four-point double-trace representation $R^\ind_{2;2}$ 
is spanned by 
\be
\{ 2~\Tr(12) \Tr(34), \qquad 2~ \Tr(13) \Tr(24),  \qquad 2~\Tr(14) \Tr(23) \}
\ee
and reduces into 
\be
{R^\ind_{2;2} \atop 3} \equals {\four \atop 1 }  \plus {\twotwo \,.\atop 2}
\ee
The one-dimensional subspace corresponding to $\four$ consists of the
symmetric sum of double-trace terms
\be
\rep{2;2}{\four} 
=  
e^\four ~ \big[ 2 \Tr(12) \Tr(34)  \big]
= 
\third \left[  
2 \Tr(12) \Tr(34)  + 2 \Tr(13) \Tr(24)  + 2 \Tr(14) \Tr(24)  
\right]
\label{twotwovertfour}
\ee
while the two-dimensional subspace corresponding to $\twotwo$ is spanned by
\be
\repi{2;2}{\twotwo} =
e_{i1} ^\twotwo \left[ 2 \Tr(12) \Tr(34)  \right]
= 
\begin{cases} 
\twothirds  \left[ 2 \Tr(12) \Tr(34)  - 2 \Tr(14) \Tr(23)  \right], & i=1\,, \\
\\
\twothirds  \left[ 2 \Tr(13) \Tr(24)  - 2 \Tr(14) \Tr(23)  \right], & i=2\,.
\end{cases}
\label{twotwoverttwotwo}
\ee
As before, 
since $\twotwo$ appears only once in $R^\ind_{2;2}$,
$\repij{2;2}{\twotwo}{2}$ 
is proportional to
$\repi{2;2}{\twotwo}$.

We apply the same approach to multiple-trace representations.
For example the irreducible subspaces of the 
induced triple-trace representation $R^\ind_{p;\,  q-p ; \, n-q}$
are spanned by 
\ba
\repij{p;q-p ;n-q}{\tau}{j}  
&\equiv &
e_{ij}^\tau 
~\big[ 
\Tr\left(  1 \cdots p \right)
\Tr\left( (p+1) \cdots q \right) 
\Tr\left( (q+1) \cdots n \right) 
\nn\\ && 
   +(-1)^n 
\Tr\left(  p \cdots 1 \right)
\Tr\left( q \cdots (p+1) \right) 
\Tr\left( n  \cdots (q+1) \right) 
\big]
\qquad
\ea
and so forth.

We will use the vectors $\repij{-}{\tau}{j}$ 
defined in this section to characterize the null spaces of
four-, five-, and six-point amplitudes in the remainder of 
this paper.

\section{Constraints on four-point amplitudes}
\setcounter{equation}{0}
\label{sec-four}

The group-theory constraints on four-point color-ordered amplitudes
at all loop orders were obtained in ref.~\cite{Naculich:2011ep} via an iterative approach.
We will not rederive these results, 
but will simply restate the $L$-loop null vectors 
(using the normalization conventions of the present paper)
and then re-express them in terms of 
irreducible representations of $S_4$,
as a warm-up for the more complicated cases to follow.

\subsection{$S_4$ decomposition of the four-point trace basis}

The four-point amplitude at any loop order may be
expanded in terms of single  and double traces of SU($N$) generators.
We express these in terms of the explicit basis\footnote{For consistency 
with the general $n$-point case (\ref{doubletraceterm}), 
we include a factor of two in the double-trace terms,
which differs from the definition in ref.~\cite{Naculich:2011ep}.
\label{factortwo}}
\ba
T_1 &=&  \Tr(1234) + \Tr(1432),
\qquad\qquad
T_4 =  2 \Tr(13) \Tr(24) , \nn\\
T_2 &=& \Tr(1243) + \Tr(1342),
\qquad\qquad
T_5=  2 \Tr(14) \Tr(23)  ,
\qquad\qquad 
\label{fourpointbasis}  \\
T_3 &=&  \Tr(1324) + \Tr(1423),
\qquad\qquad
T_6 =  2 \Tr(12) \Tr(34) . \nn
\ea
The $L$-loop four-point amplitude may then be expressed as
\be
\cA^\Ell_4 = \sum_{\la=1}^{d(L)}   A^\Ell_\la t^\Ell_\la
\label{fourpointamp}
\ee
where the $L$-loop four-point trace basis 
\be
t^\Ell_{\la+6k} = \begin{cases}
N^{L-2k} T_\la,   
& 
\la=1, 2, 3, \quad k = 0, \cdots, \lfloor {L \over 2} \rfloor  
\\
N^{L-2k-1} T_\la,   
& 
\la=4, 5, 6 , \quad k = 0, \cdots, \lfloor {L-1 \over 2} \rfloor 
\end{cases}
\label{Lloopfourpointbasis}
\ee
has dimension $d(L) = 3L + 3$.

We now decompose the four-point trace basis 
into irreducible representations of $S_4$.
The single- and double-trace terms (\ref{fourpointbasis}) 
span the representations
$R^\ind_4$ and $R^\ind_{2;2}$ respectively, which 
we recall from sec.~\ref{sec-induced} 
have the following decompositions 
\be
{R^\ind_4 \atop 3} \equals {\four \atop 1 }  \plus {\twotwo \,\,,\atop 2}
\qquad\qquad \qquad\qquad
{R^\ind_{2;2} \atop 3} \equals {\four \atop 1 }  \plus {\twotwo \,\,.\atop 2} 
\ee
The $L$-loop trace basis (\ref{Lloopfourpointbasis}) 
therefore contains $(L+1)$ copies each of $\four$ and $\twotwo$.

We can characterize the vectors $\repi{4}{\tau}$ and $\repi{2;2}{\tau}$
spanning the irreducible subspaces of $R^\ind_4$ and $R^\ind_{2;2}$
by specifying their components $v_{\la i}$ in the explicit basis 
(\ref{fourpointbasis})
\be
\repi{4}{\tau} = 
\sum_{\la=1}^3 T_\la 
\repli{4}{\tau} ,
\qquad \qquad
\repi{2;2}{\tau} = 
\sum_{\la=4}^6 T_\la 
\repli{2;2}{\tau} \,. 
\label{fourpointsubspaces}
\ee
By comparing \eqns{fourvertfour}{fourverttwotwo}
and \eqns{twotwovertfour}{twotwoverttwotwo}
with \eqns{fourpointbasis}{fourpointsubspaces} we obtain
\ba
\repl{4}{\four} 
               &=& \third \left(
                 \begin{array}{r}
                  1 \\[1mm]
                  1 \\[1mm]
                  1 \\
                 \end{array}
                 \right)\,,
\qquad\qquad
\repli{4}{\twotwo} 
	= \third	\left(
                 \begin{array}{rr}
                  0 & -1 \\[1mm]
                  1 &  1 \\[1mm]
                  -1 & 0 \\
                 \end{array}
                 \right) \,,
\label{fourpointsingletrace}
\\
\noalign{\hbox{and}}
\repl{2;2}{\four}
&=&  \third \left(
                 \begin{array}{r}
                  1 \\[1mm]
                  1 \\[1mm]
                  1 \\
                 \end{array}
                 \right)\,,
\qquad\qquad
\repli{2;2}{\twotwo}
= \twothirds
	\left(
                 \begin{array}{rr}
                  0 & 1 \\[1mm]
                  -1 & -1 \\[1mm]
                  1 & 0 \\
                 \end{array}
                 \right) \,.
\label{fourpointdoubletrace}
\ea
We will use these in the next section to decompose the four-point
null spaces into irreducible representations of $S_4$.

\subsection{Four-point null spaces}

As described in the introduction, 
the $L$-loop color space is smaller than the $L$-loop trace space,
and the null space, 
defined as the orthogonal complement of the color space,
is spanned by a set of null eigenvectors
\be
r^\Ell_{\aa} = \sum_{\la=1}^{d(L)}  r^\Ell_{\la \aa} t^\Ell_\la , 
\qquad \aa = 1, \cdots, \ncon\,.
\ee
These imply that the color-ordered amplitudes (\ref{fourpointamp})
obey a set of $\ncon$ group-theory relations
\be
\sum_\la A^\Ell_\la r^\Ell_{\la \aa} = 0, 
\qquad  \aa = 1, \cdots, \ncon\,.
\label{fourpointconstraint}
\ee

\bigsk\noindent{{\bf Even-loop four-point null space }} ($L \ge 2$) 

\noin
In ref.~\cite{Naculich:2011ep}, 
it was shown that at even-loop level (with $L \ge 2$)
the null space is four-dimensional and 
spanned by\footnote{The factor of two discrepancies with 
ref.~\cite{Naculich:2011ep} are due to 
a change in normalization; see footnote \ref{factortwo}.}
\be
r^{( 2 \ell )} = 
\begin{pmatrix} \vdots \\ 0 \\ 3 u \\ - u  \\ 0  \end{pmatrix},  \quad
\begin{pmatrix} \vdots \\ 0 \\   0 \\   2 x  \\  x \end{pmatrix}, \quad
\begin{pmatrix} \vdots \\ 0 \\   0 \\   2 y  \\  y \end{pmatrix}, \quad
\begin{pmatrix} \vdots \\ 0 \\   0 \\    0 \\  u \end{pmatrix} \,,
\qquad\qquad  \ell \ge 1
\label{evennull}
\ee
where
\be
u \equiv \begin{pmatrix} 1 \\  1 \\  1 \end{pmatrix}, \qquad
x \equiv \begin{pmatrix} 1 \\ -1 \\  0 \end{pmatrix}, \qquad 
y \equiv \begin{pmatrix} 0 \\  1 \\ -1 \end{pmatrix}. 
\ee
Comparing these with \eqns{fourpointsingletrace}{fourpointdoubletrace},
one sees that $u$ corresponds to the $\four$ representation
and $x$ and $y$ to the $\twotwo$ representation. 
The even-loop null space decomposes into 
three irreducible representations of $S_4$ spanned by
\begin{subequations} \label{eq:4pt2Lp} 
\begin{flalign}
&N^2 \rep{4}{\four} - \third  N \rep{2;2}{\four}  \\
&N  \repi{2;2}{\twotwo} - \repi{4}{\twotwo}  &i=1,2 \\
&\rep{4}{\four}  & 
\end{flalign}
\end{subequations}
Equation (\ref{eq:4pt2Lp}~c) contains only the irreducible representation
belonging to the single-trace representation $R^\ind_4$ 
and so the group-theory relation
(\ref{fourpointconstraint})
that corresponds to it 
involves only the most-subleading-color\footnote{
We refer to the amplitudes $A^\Tete$ as most-subleading-color
because they have the lowest power of $N$.}
single-trace amplitudes, namely
\be
A^\Tete (1,2,3,4) + A^\Tete (1,3,4,2) + A^\Tete (1,4,2,3) = 0
\label{fourptUfirst} 
\ee
where 
$A^ {(L,2 k)} (1,2,3,4)$ 
is the coefficient of 
$N^{L-2k} \left[ \Tr(1234)+\Tr(4321)\right]$
in $\cA^{(L)}_4$.
In general, however,
the irreducible subspaces of the null space 
are linear combinations of irreducible representations 
belonging to single- and double-trace representations. 
Thus, the corresponding group-theory constraints
relate single- and double-trace amplitudes to one another.
For example, 
eqs. (\ref{eq:4pt2Lp}~b) and  (\ref{eq:4pt2Lp}~c)
can be combined to express the most-subleading-color single-trace
amplitudes in terms of the double-trace amplitudes
\be
A^\Tete (1,2,3,4) 
=
\twothirds A^\Tetemo(1,2;3,4)
-\fourthirds A^\Tetemo(1,3;2,4)
+\twothirds A^\Tetemo(1,4;2,3)
\label{fourptDouble}
\ee
where $A^ {(L,2 k + 1)} (1,2;3,4)$ 
is the coefficient of 
$N^{L-2k-1} \left[ 2 \Tr(12)  \Tr(34) \right]$
in $\cA^{(L)}_4$.
Finally,  eq. (\ref{eq:4pt2Lp}~a) corresponds to the constraint
\ba
&& A^\Tetemt (1,2,3,4) + A^\Tetemt (1,3,4,2) + A^\Tetemt (1,4,2,3) 
\nn\\
&& - \third 
\left[A^\Tetemo (1,2; 3,4) + A^\Tetemo (1,3;4,2) + A^\Tetemo (1,4; 2,3) \right]= 0
\label{fourptUsecond} 
\ea
relating leading-color single-trace amplitudes and double-trace amplitudes.
We note that 
\eqns{fourptUfirst}{fourptUsecond},
but not \eqn{fourptDouble},
can be derived as U(1) decoupling relations (for all $\ell$).  

\bigsk\noindent{{\bf Tree-level four-point null space}}
 
\noin
At tree level, there is only one null vector, namely $r^\Zero = u$,
and so the tree-level null space is 
\begin{flalign}
&\rep{4}{\four}  & 
\end{flalign}
corresponding to the single group-theory relation (\ref{fourptUfirst})
with $\ell = 0$.

\bigsk\noindent{{\bf Odd-loop four-point null space }}   ($L \ge 3$) 

\noin
At odd-loop level (with $L \ge 3$), the null space is again four-dimensional 
and spanned by \cite{Naculich:2011ep}
\be
r^{(2\ell + 1)} = 
\begin{pmatrix} \vdots \\ 0 \\ 3 u \\ - u \\  u \\   0 \end{pmatrix} , \quad
\begin{pmatrix} \vdots \\ 0 \\ 0 \\    0 \\ 3 u \\ -  u \end{pmatrix} , \quad
\begin{pmatrix} \vdots \\ 0 \\ 0 \\    0 \\   0 \\    x \end{pmatrix} , \quad
\begin{pmatrix} \vdots \\ 0 \\   0 \\    0 \\   0 \\    y \end{pmatrix} , 
\qquad\qquad  \ell \ge 1 \,.
\label{oddnullvecs}
\ee
The odd-loop null space decomposes into three irreducible representations
of $S_4$, 
spanned by
\begin{subequations} \label{eq:4pt3Lp} 
\begin{flalign}
&N^3 \rep{4}{\four}  - \third N^2 \rep{2;2}{\four} 
+ \third N \rep{4}{\four} & \\
&N \rep{4}{\four}  - \third \rep{2;2}{\four}  & \\
&\repi{2;2}{\twotwo}   & i=1,2
\end{flalign}
\end{subequations}
Equations 
(\ref{eq:4pt3Lp}~b) and (\ref{eq:4pt3Lp}~c) 
can be combined to express the 
most-subleading-color double-trace amplitudes
in terms of single trace amplitudes
\be
A^\Teptep (1,2;3,4) =
A^\Tepte (1,2,3,4) 
+A^\Tepte (1,3,4,2) 
+A^\Tepte (1,4,2,3) \,.
\label{fourptUthird}
\ee
Finally, eq.~(\ref{eq:4pt3Lp}~a) 
gives rise to 
\ba
0 &=& 
A^\Teptemt (1,2,3,4) + A^\Teptemt (1,3,4,2) + A^\Teptemt (1,4,2,3) 
\nn\\
&& - \third 
\left[A^\Teptemo (1,2; 3,4) + A^\Teptemo (1,3;2,4) + A^\Teptemo (1,4;2,3) \right]
\nn\\
&& + \third 
\left[A^\Tepte (1,2,3,4) + A^\Tepte (1,3,4,2) + A^\Tepte (1,4,2,3) \right]\,.
\label{fourptOdd}
\ea
We note that \eqn{fourptUthird}
follows from U(1) decoupling (for all $\ell$)
but \eqn{fourptOdd} does not. 

\bigsk\noindent{{\bf One-loop four-point null space }} 

\noin
The one-loop null space is spanned by the last three vectors of 
\eqn{oddnullvecs}
and breaks into two irreducible subspaces spanned by
\begin{subequations} 
\begin{flalign}
&N \rep{4}{\four}  - \third \rep{2;2}{\four}  & \\
&\repi{2;2}{\twotwo}   & i=1,2
\end{flalign}
\end{subequations}
These give rise to the constraint (\ref{fourptUthird}) with $\ell=0$,
which follows from U(1) decoupling.

\section{Constraints on five-point amplitudes}
\setcounter{equation}{0}
\label{sec-five}

The group-theory constraints on five-point color-ordered amplitudes
at all loop orders were obtained in ref.~\cite{Edison:2011ta}.
These were expressed in terms of a complete set of null vectors
by giving their components with respect to an explicit basis.
In this section, we show how these results can be expressed 
more compactly by writing the null space in terms
of irreducible representations of $S_5$.

\subsection{$S_5$ decomposition of the five-point trace basis} 

Five-point amplitudes at any loop order can be expanded in a basis
including single and double traces of SU($N$) generators. 
As we described in sec.~\ref{sec-trace}, 
the twelve independent single-trace terms of the form
\be
\Tr( 12345) - \Tr( 54321)
\label{fivepointsingletrace}
\ee
are labeled $T_\la$  with $\la = 1, \cdots, 12$,
and the ten independent double-trace terms of the form
\be
\Tr(12)  \Tr(345) -  \Tr(21) \Tr(543)
\label{fivepointdoubletrace}
\ee
are labeled $ T_\la $ with $\la = 13, \cdots, 22$.
An explicit ordering for the basis $\{ T_\la \}$ 
is given in appendix \ref{app-five},
but the results presented in this section will
be independent of the specific choice of basis. 

The $L$-loop five-point amplitude can be expressed as 
\be
\cA^\Ell_5 = \sum_{\la=1}^{d(L)}   A^\Ell_\la t^\Ell_\la
\ee
where the $L$-loop  five-point trace basis 
\be
t^\Ell_{\la+ 22k} 
= \begin{cases}  
N^{L - 2k} \, T_\la  \,, 
&  
\la = 1, \cdots, 12, 
 ~\qquad k = 0, \cdots, \lfloor \frac{L}{2}  \rfloor
\\ 
N^{L-2k-1} \, T_\la  \,, 
& 
\la = 13, \cdots, 22,
\qquad k = 0, \cdots, \lfloor \frac{L-1}{2}  \rfloor 
\end{cases}
\ee
has dimension 
\be
d(L) = 
\begin{cases}
12 + 11 L,  &\hbox{ for  $L$ even,} \\ 
11 + 11 L,  &\hbox{ for  $L$ odd.}
\end{cases}
\ee

Now we decompose the trace basis into irreducible representations of $S_5$.
The induced representations
$R^\ind_5$ and $R^\ind_{2;3}$,
spanned by the single-trace
(\ref{fivepointsingletrace})
and double-trace (\ref{fivepointdoubletrace}) terms respectively, 
have the following decompositions 
\be
{R^\ind_5 \atop 12} \equals {2 ~\threeoneone \atop 2 \cdot 6 }
\qquad\qquad \qquad\qquad
{R^\ind_{2;3} \atop 10} \equals {\threeoneone \atop 6 }  
\plus {\twooneoneone \,\,.\atop 4}
\label{fivepointdecomp}
\ee
As described in sec.~\ref{sec-regular},
to define vectors in the trace space that 
span the irreducible subspaces  $\threeoneone$ and $\twooneoneone$, 
we define standard tableaux $\tau_i$ for each of these representations
\be
\tau_1 = 
{\def\lr#1{\multicolumn{1}{|@{\hspace{.6ex}}c@{\hspace{.6ex}}|}{\raisebox{-.3ex}{$#1$}}}
\raisebox{-.6ex}{$\begin{array}[b]{ccc}
\cline{1-1}\cline{2-2}\cline{3-3}
\lr{1}&\lr{2}&\lr{3}\\
\cline{1-1}\cline{2-2}\cline{3-3}
\lr{4}\\
\cline{1-1}
\lr{5}\\
\cline{1-1}
\end{array}$} }
\quad
\tau_2 = 
{\def\lr#1{\multicolumn{1}{|@{\hspace{.6ex}}c@{\hspace{.6ex}}|}{\raisebox{-.3ex}{$#1$}}}
\raisebox{-.6ex}{$\begin{array}[b]{ccc}
\cline{1-1}\cline{2-2}\cline{3-3}
\lr{1}&\lr{2}&\lr{4}\\
\cline{1-1}\cline{2-2}\cline{3-3}
\lr{3}\\
\cline{1-1}
\lr{5}\\
\cline{1-1}
\end{array}$}}
\quad
\tau_3 = 
{\def\lr#1{\multicolumn{1}{|@{\hspace{.6ex}}c@{\hspace{.6ex}}|}{\raisebox{-.3ex}{$#1$}}}
\raisebox{-.6ex}{$\begin{array}[b]{ccc}
\cline{1-1}\cline{2-2}\cline{3-3}
\lr{1}&\lr{2}&\lr{5}\\
\cline{1-1}\cline{2-2}\cline{3-3}
\lr{3}\\
\cline{1-1}
\lr{4}\\
\cline{1-1}
\end{array}$}
}
\quad
\tau_4 = 
{\def\lr#1{\multicolumn{1}{|@{\hspace{.6ex}}c@{\hspace{.6ex}}|}{\raisebox{-.3ex}{$#1$}}}
\raisebox{-.6ex}{$\begin{array}[b]{ccc}
\cline{1-1}\cline{2-2}\cline{3-3}
\lr{1}&\lr{3}&\lr{4}\\
\cline{1-1}\cline{2-2}\cline{3-3}
\lr{2}\\
\cline{1-1}
\lr{5}\\
\cline{1-1}
\end{array}$}
}
\quad
\tau_5 = 
{\def\lr#1{\multicolumn{1}{|@{\hspace{.6ex}}c@{\hspace{.6ex}}|}{\raisebox{-.3ex}{$#1$}}}
\raisebox{-.6ex}{$\begin{array}[b]{ccc}
\cline{1-1}\cline{2-2}\cline{3-3}
\lr{1}&\lr{3}&\lr{5}\\
\cline{1-1}\cline{2-2}\cline{3-3}
\lr{2}\\
\cline{1-1}
\lr{4}\\
\cline{1-1}
\end{array}$}
}
\quad
\tau_6 = 
{\def\lr#1{\multicolumn{1}{|@{\hspace{.6ex}}c@{\hspace{.6ex}}|}{\raisebox{-.3ex}{$#1$}}}
\raisebox{-.6ex}{$\begin{array}[b]{ccc}
\cline{1-1}\cline{2-2}\cline{3-3}
\lr{1}&\lr{4}&\lr{5}\\
\cline{1-1}\cline{2-2}\cline{3-3}
\lr{2}\\
\cline{1-1}
\lr{3}\\
\cline{1-1}
\end{array}$}
}
\ee
\be
\tau_1 = 
{\def\lr#1{\multicolumn{1}{|@{\hspace{.6ex}}c@{\hspace{.6ex}}|}{\raisebox{-.3ex}{$#1$}}}
\raisebox{-.6ex}{$\begin{array}[b]{cc}
\cline{1-1}\cline{2-2}
\lr{1}&\lr{2}\\
\cline{1-1}\cline{2-2}
\lr{3}\\
\cline{1-1}
\lr{4}\\
\cline{1-1}
\lr{5}\\
\cline{1-1}
\end{array}$}
}
\quad
\tau_2 = 
{\def\lr#1{\multicolumn{1}{|@{\hspace{.6ex}}c@{\hspace{.6ex}}|}{\raisebox{-.3ex}{$#1$}}}
\raisebox{-.6ex}{$\begin{array}[b]{cc}
\cline{1-1}\cline{2-2}
\lr{1}&\lr{3}\\
\cline{1-1}\cline{2-2}
\lr{2}\\
\cline{1-1}
\lr{4}\\
\cline{1-1}
\lr{5}\\
\cline{1-1}
\end{array}$}
}
\quad
\tau_3 = 
{\def\lr#1{\multicolumn{1}{|@{\hspace{.6ex}}c@{\hspace{.6ex}}|}{\raisebox{-.3ex}{$#1$}}}
\raisebox{-.6ex}{$\begin{array}[b]{cc}
\cline{1-1}\cline{2-2}
\lr{1}&\lr{4}\\
\cline{1-1}\cline{2-2}
\lr{2}\\
\cline{1-1}
\lr{3}\\
\cline{1-1}
\lr{5}\\
\cline{1-1}
\end{array}$}
}
\quad
\tau_4 = 
{\def\lr#1{\multicolumn{1}{|@{\hspace{.6ex}}c@{\hspace{.6ex}}|}{\raisebox{-.3ex}{$#1$}}}
\raisebox{-.6ex}{$\begin{array}[b]{cc}
\cline{1-1}\cline{2-2}
\lr{1}&\lr{5}\\
\cline{1-1}\cline{2-2}
\lr{2}\\
\cline{1-1}
\lr{3}\\
\cline{1-1}
\lr{4}\\
\cline{1-1}
\end{array}$}
}
\ee
From these, we construct the projection operators $e_{ij}^\tau$ using \eqn{young}.
Acting with the projection operators on single- and double-trace
representations, we obtain the vectors
\ba
\repij{5}{\tau}{j} 
&=&
e_{ij}^\tau \left[  \Tr(12345) - \Tr(54321) \right] \,,
\label{fivepointsingletracedecomp}
\\
\repij{2;3}{\tau}{j} 
&=&
e_{ij}^\tau \left[  \Tr(12) \Tr(345) - \Tr(21) \Tr(543) \right] 
\label{fivepointdoubletracedecomp}
\ea
that span the irreducible subspaces.
Needless to say, all of this is done using a symbolic
manipulation program.
In appendix \ref{app-five},
we write down the components of 
$\repij{5}{\tau}{j} $ and $\repij{2;3}{\tau}{j} $
in an explicit basis.

\subsection{Five-point null spaces}

In ref.~\cite{Edison:2011ta},
it was shown that the five-point $L$-loop null space
is spanned by a set of null vectors
\be
r^\Ell_{\aa} = \sum_{\la=1}^{d(L)}  r^\Ell_{\la \aa} t^\Ell_\la , 
\qquad \aa = 1, \cdots,
\begin{cases}
6,   & L=0 \\
10,  & {\rm odd~} L  \\
12,  & {\rm even~} L \ge 2\,.\\
\end{cases}
\ee
We will not rederive these null vectors, but simply
restate them in terms of irreducible spaces.

\bigsk\noindent{{\bf Tree-level five-point null space } } 

\noin
The 6 tree-level null vectors belong to a representation of $S_5$ 
contained within the tree-level trace space  $R^\ind_5$ 
and so must correspond to some linear combination of the two copies of
$\threeoneone$ contained therein (cf. \eqn{fivepointdecomp}):
\be
\kappa_1 \repij{5}{\threeoneone}{1}  + \kappa_2  \repij{5}{\threeoneone}{2} \,.
\ee
By writing these vectors in an explicit basis using \eqn{explicitsixdimrep}
and then acting with $M^\Zero$ (cf. eq. (3.9) of ref.~\cite{Edison:2011ta}),
we find a result proportional to $\kappa_1 - 3 \kappa_2 $.
Hence the tree-level five-point null space is spanned by 
\begin{flalign}
\label{eq:5ptTree} 
&3 \repij{5}{\threeoneone}{1}  + \repij{5}{\threeoneone}{2} 
& i = 1, \cdots, 6 
\end{flalign}
The 6 null vectors (\ref{eq:5ptTree}) give rise to the relations 
among tree-level amplitudes \cite{Kleiss:1988ne,DelDuca:1999rs}
\be
A^\Zerozero (1, \setal, 5, \setbe )
= (-1)^{n_\beta} \sum_{\sig \in OP \setal \setbeT } 
A^\Zerozero (1,\sig, 5)
\label{fiveptUfirst}
\ee
where $A^\Elltk (1,2,\cdots,n)$ 
is the coefficient of 
$ N^{L-2k} \left[ \Tr  (  12 \cdots n ) + (-1)^n \Tr  (  n \cdots 21 ) \right]$
in $\cA^{(L)}_n$,
and $\setal$ and $\setbe$ denote complementary subsets of
$\{ 2, 3, \cdots, n-1 \}$,
with $n_\beta$ the number of elements of $\{\beta\}$.
Here $OP\setal\setbeT$ denotes the set of 
{\it ordered permutations} (mergings) of $\setal$ and $\setbeT$,
those that preserve the order of $\setal$ and $\setbeT$ inside $\sig$,
where $\setbeT$ is $\setbe$ with the ordering reversed.
\Eqn{fiveptUfirst} can be derived from U(1) decoupling.

\bigsk\noindent{{\bf Odd-loop five-point null space}}

\noin
At odd-loop level,
the five-point null space is 10-dimensional,
and breaks into two irreducible subspaces, spanned by
\begin{subequations} \label{eq:5pt1Lp}
\begin{flalign}
& 3 N \repij{5}{\threeoneone}{1}  + N \repij{5}{\threeoneone}{2} 
- \half   \repi{2;3}{\threeoneone} 
&i = 1, \cdots, 6 \\
& \repi{2;3}{\twooneoneone} 
& i = 1, \cdots,  4
\end{flalign}
\end{subequations}
Since the null space involves both irreducible representations
$\threeoneone$ and $\twooneoneone$ of $R^\ind_{2;3}$, 
the null vectors are sufficient to express the
most-subleading-color double-trace amplitudes in terms of 
single-trace amplitudes
\be
A^\Teptep(1,2; ~3,4,5 ) 
= \sum_{\sig \in COP \{2,1\} \{3,4,5\} } A^\Tepte(\sig)
\label{fiveptUsecond}
\ee
where 
$A^\Elltkp (1,\cdots,p ; ~ p+1,\cdots,n)$ 
is the coefficient of 
$N^{L-2k-1} \big[ \Tr(1\cdots p) \Tr((p+1) \cdots n) + $
$(-1)^n \Tr(p \cdots 1) \Tr(n\cdots (p+1))\big]$
in $\cA^{(L)}_n$,
and $COP\setal\setbe$ is the set of {\it cyclically-ordered permutations},
i.e., permutations of $\{1, \cdots, n \}$  (with $n$ held fixed)
that preserve the cyclic ordering of $\setal$ and $\setbe$.
\Eqn{fiveptUsecond} can be derived from U(1) decoupling (for all $\ell$),
and extends the one-loop results of ref.~\cite{Bern:1990ux}
to all odd-loop orders.

\bigsk\noindent{{\bf Even-loop five-point null space} } ($L \ge 2$)

\noin
At even-loop level (with $L \ge 2$) the 
null space is 12-dimensional and 
decomposes into two irreducible subspaces spanned by 
\begin{subequations} \label{eq:5pt2Lp}
\begin{flalign}
&
3 N^2 \repij{5}{\threeoneone}{1}  +  N^2 \repij{5}{\threeoneone}{2} 
- \half  N  \repi{2;3}{\threeoneone}
+ \repij{5}{\threeoneone}{1} & i = 1, \cdots, 6 \\
& 3 \repij{5}{\threeoneone}{1}  +  \repij{5}{\threeoneone}{2} 
&i = 1, \cdots, 6 
\end{flalign}
\end{subequations} 
Equation (\ref{eq:5pt2Lp}b) gives rise to 6 relations
among most-subleading-color single-trace amplitudes
\be
A^\Tete (1, \setal, 5, \setbe )
= (-1)^{n_\beta} \sum_{\sig \in OP \setal \setbeT } 
A^\Tete (1,\sig, 5) \,.
\label{fiveptUthird}
\ee
These relations can be derived (for all $\ell$) from U(1) decoupling.

Equations (\ref{eq:5pt2Lp}a) 
and (\ref{eq:5pt2Lp}b) 
can be combined to express each of the most-subleading-color 
single-trace amplitudes $A^\Tete$ 
in terms of double-trace amplitudes $A^\Tetemo$ and single-trace amplitudes
$A^\Tetemt$.
Explicit expressions for these were given in ref.~\cite{Edison:2011ta}.
These relations, however, cannot be derived using U(1) decoupling.

\section{Constraints on six-point amplitudes}
\setcounter{equation}{0}
\label{sec-six}

In this section, we derive the group-theory relations
obeyed by six-point color-ordered amplitudes to all loop orders.
First we describe the decomposition of the six-point trace basis
into irreducible representations of $S_6$.
Then we derive the $L$-loop six-point null space 
using the iterative approach of refs.~\cite{Naculich:2011ep,Edison:2011ta},
and express the results in terms of irreducible subspaces of $S_6$. 

\subsection{$S_6$ decomposition of the six-point trace basis}
\label{sec-sixtrace}

Six-point amplitudes at any loop order can be expanded in a basis
that includes single, double, and triple traces of SU($N$) generators. 
We label the single-, double-, and triple-trace terms 
as $T_\la$ with $\la=1, \cdots, 140$, as described in sec.~\ref{sec-trace}.
The sixty independent single-trace terms of the form
\be
\Tr( 123456) + \Tr( 654321)
\ee
are labeled $T_\la$  with $\la = 1, \cdots, 60$.
The precise ordering of the terms is unimportant 
since we will present our results in a form independent of it.
These single-trace terms span the representation $R^\ind_6$ 
which decomposes into
\be
{R^\ind_6 \atop 60}
\equals
{\six \atop 1}  \plus 
{2 ~\fourtwo \atop 2 \cdot 9}  \plus
{\threetwoone \atop 16    }  \plus
{2 ~\twotwotwo \atop 2 \cdot 5}  \plus
{\threeoneoneone \atop 10   }  \plus
{\twooneoneoneone  \,\,.  \atop 5  }  
\label{sixdecomp}
\ee
At loop level, we also require forty-five independent
double-trace terms of the form
\be
\Tr(12)  \Tr(3456) +  \Tr(21) \Tr(6543)
\ee
which we label $ T_\la $ with $\la = 61, \cdots, 105$.
These span $R^\ind_{2;4}$ which decomposes into 
\be
{R^\ind_{2;4} \atop 45}
\equals
{\six \atop 1}  \plus 
{\fiveone \atop 5}  \plus 
{2 ~\fourtwo \atop 2\cdot 9}  \plus
{\threetwoone \atop 16}  \plus
{\twotwotwo\,\,.   \atop 5}  
\label{twofourdecomp}
\ee
There are also twenty independent double-trace terms of the form 
\be
\Tr(123)  \Tr(456) + \Tr(321) \Tr(654) 
\ee
labeled $ T_\la $ with $\la = 106, \cdots, 125$.
These span $R^\ind_{3;3}$, which decomposes into 
\be
{R^\ind_{3;3} \atop 20}
\equals
{\six \atop 1}  \plus 
{\fourtwo \atop 9}  \plus 
{\twotwotwo \atop 5}  \plus
{\twooneoneoneone\,\,.    \atop 5}  
\label{threethreedecomp}
\ee
Finally, at two loops and above, there are fifteen independent
triple-trace terms of the form 
\be 
\Tr(12) \Tr(34) \Tr(56) + \Tr(21) \Tr(43) \Tr(65)
\qquad =  \qquad
2 \Tr(12) \Tr(34) \Tr(56)
\ee 
labeled $ T_\la $ with $\la = 126, \cdots, 140$.
These span $R^\ind_{2;2;2}$, which decomposes into 
\be
{R^\ind_{2;2;2} \atop 15}
\equals
{\six \atop 1}  \plus 
{\fourtwo \atop 9}  \plus 
{\twotwotwo \,\,.  \atop 5} 
\label{twotwotwodecomp}
\ee
As before, by constructing projection operators $e_{ij}^\tau$ 
for each of the tableaux appearing in the induced representations,
one can project those representations onto the vectors 
\be
\repij{6}{\tau}{j}, \qquad
\repij{2;4}{\tau}{j}, \qquad
\repij{3;3}{\tau}{j}, \qquad
\repij{2;2;2}{\tau}{j}
\ee
that span the irreducible subspaces of 
$R^\ind_6$,
$R^\ind_{2;4}$,
$R^\ind_{3;3}$, and 
$R^\ind_{2;2;2}$ respectively.
For obvious reasons we will not give explicit
expressions for these.
If the label $j$ is omitted, one should presume $j=1$,
except in the following cases
\be
\repi{6}{\threeoneoneone} \equiv \repij{6}{\threeoneoneone}{2}, \quad
\repi{3;3}{\twotwotwo} \equiv \repij{3;3}{\twotwotwo}{5}, \quad
\repi{3;3}{\twooneoneoneone} \equiv \repij{3;3}{\twooneoneoneone}{3},
\ee
because the vectors for smaller values of $j$ vanish in these cases.

The $L$-loop six-point amplitude can be expanded as
\be
\cA^\Ell_6  = \sum_{\la=1}^{d(L)}   A^\Ell_\la t^\Ell_\la
\label{sixpointamplitude}
\ee
where the $L$-loop six-point trace basis 
\be
t^\Ell_{\la+ 140k} 
= \begin{cases}  
N^{L - 2k} \, T_\la  \,, 
&  
\la = 1, \cdots, 60, 
~~~~\qquad k = 0, \cdots, \lfloor \frac{L}{2}  \rfloor
\\ 
N^{L-2k-1} \, T_\la  \,, 
& 
\la = 61, \cdots, 125,
~~\qquad k = 0, \cdots, \lfloor \frac{L-1}{2}  \rfloor 
\\
N^{L-2k-2} \, T_\la  \,, 
& 
\la = 126, \cdots, 140,
~\qquad k = 0, \cdots, \lfloor \frac{L-2}{2}  \rfloor 
\end{cases}
\label{Lloopsixpointtracebasis}
\ee
has dimension 
\be
d(L) = \begin{cases}   
60 + 70 L, & \hbox{ for  $L$ even,} \\
55 + 70 L, & \hbox{ for  $L$ odd.}
\end{cases}
\label{dimtracespace}
\ee

\subsection{Six-point null spaces}

Having defined the $L$-loop six-point trace space,
we now turn to the determination 
of a complete set of $L$-loop six-point color factors 
\be
c^\Ell_i = \sum_{\la = 1}^{d(L)}  M^\Ell_{ i \la } t^\Ell_\la  
\label{sixpointcolorfactor}
\ee 
which span a subspace of the trace space (the color space).
The null space, which is the orthogonal complement\footnote{
Orthogonality is defined with respect to the inner product 
$(t^\Ell_\la, t^\Ell_{\la'} ) = \delta_{\la \la'}$.
}
of the color space, is spanned by the null vectors
\be
r^\Ell_\aa = \sum_{\la = 1}^{d(L)}  r^\Ell_{\la \aa} t^\Ell_\la, 
\qquad \hbox{where} \qquad
\sum_{\la = 1}^{d(L)}  M^\Ell_{ i \la } r^\Ell_{\la \aa}   = 0 \,.
\ee
A complete set of null vectors then determines the group-theory relations
\be
\sum_\la A^\Ell_\la r^\Ell_{\la \aa} = 0, 
\qquad  \aa = 1, \cdots, \ncon
\ee
satisfied by the six-point color-ordered amplitudes 
defined in \eqn{sixpointamplitude}.

\bigsk\noindent{{\bf Tree-level six-point null space}} 

\noin
We begin with the tree-level six-point color space, spanned by 
the 24 independent color factors \cite{DelDuca:1999ha,DelDuca:1999rs}
\be
\f^{a_1 a_j b}
\f^{b a_k c} 
\f^{c a_l d} 
\f^{d a_m a_6} 
\label{ddm}
\ee
where $\{j,k,l,m\}$ runs over all permutations of $\{2,3,4,5\}$. 
We expand these in the tree-level trace basis $\{ t^\Zero_\la \}$ 
to find 
$M^\Zero_{i \la}$
and then solve for its 36 null eigenvectors $r^\Zero_{\la\aa}$.
They span a null space that decomposes into
the following irreducible representations of $S_6$:
\begin{subequations} \label{eq:6ptTree}
\begin{flalign}
&\rep{6}{\six}&  \\
&\repij{6}{\fourtwo}{1} +\fourth \repij{6}{\fourtwo}{2} &i=1,\cdots,9  \\
&\repi{6}{\threetwoone} &i=1,\cdots,16 \\
&\repij{6}{\twotwotwo}{1}+\repij{6}{\twotwotwo}{2} & i=1,\cdots,5 \\
&\repi{6}{\twooneoneoneone} & i=1,\cdots,5
\end{flalign}
\end{subequations}
These null vectors give rise to the 36 relations among tree-level 
amplitudes \cite{Kleiss:1988ne,DelDuca:1999rs}
\be
A^\Zerozero (1, \setal, 6, \setbe )
= (-1)^{n_\beta} \sum_{\sig \in OP \setal \setbeT } 
A^\Zerozero (1,\sig, 6) \,.
\label{sixptUfirst}
\ee
\Eqn{sixptUfirst}
can be derived from U(1) decoupling \cite{Kleiss:1988ne}.

Next we turn to the iterative construction of the loop-level color basis.
An $(L+1)$-loop color diagram may be obtained from an $L$-loop color diagram 
by attaching a rung between two of its external legs, $j$ and $k$. 
First, we describe the effect of adding a rung 
on the trace basis $\{T_\la\}$ defined in the previous section.
Choose an arbitrary trace term $T_\la$, 
contract it with $\f^{a_j a'_j b} \f^{b a'_k a_k}$,
and simplify using \eqn{structure} together with 
the SU($N$) relations 
\ba
\Tr(P T^a) \Tr(Q T^a) &=& \Tr (PQ) - {1 \over N} \Tr(P) \Tr(Q) \,,
\nn\\
\Tr(P T^a Q T^a) &=& \Tr (P) \Tr(Q) - {1 \over N} \Tr(P Q) \,.
\label{sunrelations}
\ea
This procedure yields a linear combination of trace terms
\be
T_\la \longrightarrow \sum_{\ka=1}^{140} G_{\la\ka} T_\ka  \,.
\ee
Attaching a rung to single-trace terms yields both single-
and double-trace terms; 
attaching a rung to a triple-trace term yields double-
and triple-trace terms, 
and attaching a rung to a double-trace terms yields all three.
Hence the $140 \times 140$ matrix $G$ has the block form
\be 
G = 
\begin{pmatrix} 
N \Aa & \Ab & \Ac & 0 \\ 
\Ad & N \Ae & \Af & \Ag \\
\Ah & \Ai & N \Aj & \Ak \\
0 & \Al & \Am &  N \An 
\end{pmatrix}
\label{defG}
\ee
with blocks of size 60, 45, 20, and 15 respectively,
where the $N$ dependence is made explicit.
The matrix elements depend, of course, on the choice of legs 
$j$ and $k$ between which the rung extends.

Now using \eqn{defG}, we can determine the effect 
of  contracting an arbitrary element $t^\Ell_\la$ 
of the $L$-loop trace basis (\ref{Lloopsixpointtracebasis})
with $\f^{a_j a'_j b} \f^{b a'_k a_k}$.
The result is a linear combination of
elements of the $(L+1)$-loop trace basis 
\be
t^\Ell_\la \to \sum_{\ka=1}^{d(L+1)} g^\Ell_{\la\ka} t^\Ellplus_\ka 
\label{procedure}
\ee
where $g^\Ell$ is the $d(L) \times d(L+1)$ matrix 
\be
g^\Ell = \left( \begin{array}{ccccccccc}
	\Aa & \Ab & \Ac & 0   & 0   & 0   & 0   &  \hdots \\
	0   & \Ae & \Af & \Ag & \Ad & 0   & 0   &  \hdots \\
	0   & \Ai & \Aj & \Ak & \Ah & 0   & 0   &  \hdots \\
	0   & 0   & 0   & \An &  0  & \Al & \Am  & \hdots \\
	0   & 0   & 0   &  0  & \Aa & \Ab & \Ac  & \hdots \\
	\vdots & \vdots & \vdots & \vdots & \vdots & 
        \vdots & \vdots & \ddots \\
	\end{array} \right) \,.
\label{contract}
\ee
The pattern repeats 
starting from $g^\Zero = ( a ~~ b~~ c )$,
with each increment of $L$ adding two extra rows and columns of blocks.

Now we make the assumption that we can construct 
the complete $(L+1)$-loop color space by attaching rungs 
between the legs of $L$-loop color factors in all possible ways.
We act on the $L$-loop color basis (\ref{sixpointcolorfactor})
with \eqn{procedure} 
to obtain the $(L+1)$-loop color basis
\be
\sum_{\la=1}^{d(L)}  \sum_{\ka=1}^{d(L+1)} 
M^\Ell_{ i \la } g^\Ell_{\la\ka} t^\Ellplus_\ka  \,.
\ee
The $(L+1)$-loop null vectors $r^\Ellplus_\aa$ 
must then satisfy
\be
\sum_{\la=1}^{d(L)}   \sum_{\ka=1}^{d(L+1)} M^\Ell_{i  \la} g^\Ell_{\la \ka} r^\Ellplus_{\ka \aa} = 0 \,.
\label{recursion}
\ee
\Eqn{recursion}
must be satisfied for all possible choices of $j$ and $k$ used
to construct   $g^\Ell$.
We solve \eqn{recursion} 
to obtain a complete set of $(L+1)$-loop null vectors
$\{ r^\Ellplus_\aa \}$.
Finally, we construct $M^\Ellplus_i$ 
as the set of vectors orthogonal to $r^\Ellplus_\aa$,
and begin the process again.

\bigsk\noindent{{\bf One-loop six-point null space}}

\noin
Using this procedure, 
we find 65 null vectors at the one-loop level. 
The null space 
can be decomposed into $S_6$ subspaces spanned by
\begin{subequations} \label{eq:6pt1Lp}
\begin{flalign}
& N \rep{6}{\six}-\nicefrac{1}{20} \rep{2;4}{\six}  \\
& N \repij{6}{\fourtwo}{1} +\fourth N \repij{6}{\fourtwo}{2}-\nicefrac{1}{8} \repij{2;4}{\fourtwo}{1}  & i=1,\cdots,9   \\
& N \repi{6}{\threetwoone}-\fourth \repi{2;4}{\threetwoone} & i=1,\cdots,16  \\
& N \repij{6}{\twotwotwo}{1}+N \repij{6}{\twotwotwo}{2} +\half \repi{2;4}{\twotwotwo} & i=1,\cdots, 5  \\
& N \repi{6}{\twooneoneoneone}-\nicefrac{1}{6} \repi{3;3}{\twooneoneoneone}  & i=1,\cdots,5  \\
& \rep{2;4}{\six}+\twothirds \rep{3;3}{\six}  \\
& \repi{2;4}{\fiveone} & i=1,\cdots,5  \\
& \repij{2;4}{\fourtwo}{1} +\half \repij{2;4}{\fourtwo}{2}  & i=1,\cdots,9   \\
& \repi{2;4}{\twotwotwo}+\nicefrac{1}{6} \repi{3;3}{\twotwotwo} & i=1,\cdots,5  \\
&\repi{3;3}{\fourtwo} & i=1,\cdots,9
\end{flalign}
\end{subequations}
Since all of the irreducible representations in 
the double-trace representations 
(\ref{twofourdecomp}) and (\ref{threethreedecomp})
appear in the null space,
the corresponding constraints are sufficient to express all 65 
of the double-trace amplitudes in terms of single-trace amplitudes.
Specifically,
eqns.~(\ref{eq:6pt1Lp}~a,b,c,d,g,h) can be combined to solve for 
the 45 independent 2;4 double-trace amplitudes
\be
A^\Oneone(1,2; ~ 3,4,5,6 ) 
= \sum_{\sig \in COP \{2,1\} \{3,4,5,6\} } A^\Onezero(\sig) \,.
\label{sixptUsecond}
\ee
These relations were deduced in ref.~\cite{Bern:1990ux}
using U(1) decoupling.
Equations (\ref{eq:6pt1Lp}~e,f,i,j) together with eqs.~(\ref{eq:6pt1Lp}~a,d) 
yield the 20 independent 3;3 double-trace amplitudes
\be
A^\Oneone(1,2, 3; ~4,5,6 ) 
= ~(-1)~ \sum_{\sig \in COP \{3,2,1\} \{4,5,6\} } A^\Onezero(\sig) \,.
\label{sixptDouble} 
\ee
These relations were obtained in ref.~\cite{Bern:1994zx}
using string-theoretic methods,
and later proved field-theoretically in ref.~\cite{DelDuca:1999rs}.

Because U(1) decoupling relations only involve the
{\it symmetrized} 3;3 amplitudes 
\be
S_{123}^\Elltkp \equiv  A^\Elltkp (1,2,3; ~4,5,6) + A^\Elltkp (1,2,3; ~4,6,5)
\label{sym33}
\ee
they are not sufficient to derive \eqn{sixptDouble},
but can be used to obtain \cite{Bern:1990ux}
\be
S_{123}^\Oneone 
= ~(-1)~ \sum_{\sig \in COP \{3,2,1\} } A^\Onezero(\sig) \,.
\label{sixptUthird}
\ee
Since the 10 symmetrized 3;3 amplitudes are contained
in the irreducible representations
$\repi{3;3}{\six}$ and $\repi{3;3}{\fourtwo}$,
\eqn{sixptUthird}
follows from eqs.~(\ref{eq:6pt1Lp}~a,f,j) alone.
As we will see below, the U(1) decoupling relations (\ref{sixptUsecond}) 
and (\ref{sixptUthird})
continue to hold at higher (odd) loops, whereas \eqn{sixptDouble} 
is modified.

\bigsk\noindent{{\bf Two-loop six-point null space}}

\noin
At the two-loop level, we find 80 null vectors, which can be 
decomposed into irreducible subspaces  spanned by
\begin{subequations} \label{eq:6pt2Lp}
\begin{flalign}
& N^2 \rep{6}{\six}-\nicefrac{1}{20} N\rep{2;4}{\six}-\nicefrac{1}{  30} \rep{2;2;2}{\six} \\
& \nn \\
& N^2 \repij{6}{\fourtwo}{1} +\fourth N^2 \repij{6}{\fourtwo}{2}-\nicefrac{1}{4} N \repij{2;4}{\fourtwo}{1} \nn \\
& \qquad-\nicefrac{1}{16} N \repij{2;4}{\fourtwo}{2}+ \fourth\repij{6}{\fourtwo}{1}& i=1,\cdots,9 \\
 & \nn \\
& N^2 \repij{6}{\twotwotwo}{1}+N^2 \repij{6}{\twotwotwo}{2} +\nicefrac{1}{10} N \repi{2;4}{\twotwotwo}  \nn \\
& \qquad -\nicefrac{1}{15} N \repi{3;3}{\twotwotwo} -\nicefrac{2}{15} \repij{6}{\twotwotwo}{1}& i=1,\cdots,5 \\
& \nn \\
& N \rep{2;4}{\six}+\twothirds N \rep{3;3}{\six} -\nicefrac{1}{   3} \rep{2;2;2}{\six} \\
& N \repi{2;4}{\fiveone} & i=1,\cdots,5 \\
& N \repij{2;4}{\fourtwo}{1} +\half N\repij{2;4}{\fourtwo}{2} -\nicefrac{1}{   4} \repi{2;2;2}{\fourtwo} & i=1,\cdots,9  \\
 & N \repi{3;3}{\fourtwo} +\nicefrac{3}{   8} \repi{2;2;2}{\fourtwo}& i=1,\cdots,9 \\
& \repi{2;2;2}{\twotwotwo}& i=1,\cdots,5 \\
& \rep{6}{\six}  \\
& \repij{6}{\fourtwo}{1} +\fourth \repij{6}{\fourtwo}{2} & i=1,\cdots,9   \\
& \repi{6}{\threetwoone} & i=1,\cdots,16 \\
& \repij{6}{\twotwotwo}{1}+\repij{6}{\twotwotwo}{2} & i=1,\cdots,5 \\
& \repi{6}{\twooneoneoneone} & i=1,\cdots,5
\end{flalign}
\end{subequations}
The two-loop null vectors (\ref{eq:6pt2Lp}~i-m)
exactly parallel the tree-level null vectors (\ref{eq:6ptTree})
and therefore lead to 36 independent relations 
(previously obtained by Feng et al. \cite{Feng:2011fja})
\be
A^\Twotwo (1, \setal, 6, \setbe )
= (-1)^{n_\beta} \sum_{\sig \in OP \setal \setbeT } 
A^\Twotwo (1,\sig, 6)
\label{SCiden}
\ee
among the subleading-color single-trace amplitudes.
The relations (\ref{SCiden}) can be derived using U(1) decoupling.
Equations (\ref{eq:6pt2Lp}~b,c) can be used to obtain 14 additional constraints
(not derivable from U(1) decoupling) relating the 
subleading-color single-trace amplitudes
to leading-color single-trace amplitudes
as well as double- and triple-trace amplitudes.
Unlike the four- and five-point cases, however, one cannot use the 
set of null vectors (\ref{eq:6pt2Lp}) to solve for all 60 
subleading-color single-trace amplitudes $A^\Twotwo$ in terms of other amplitudes.
This is most easily seen by observing that $\repi{6}{\threeoneoneone}$
is absent from the null space (\ref{eq:6pt2Lp}).

Since the two-loop null vectors (\ref{eq:6pt2Lp}~a,f,h) 
contain all of the irreducible representations in 
the triple-trace representation (\ref{twotwotwodecomp}),
they are sufficient to solve for the 15 triple-trace amplitudes
in terms of double-trace amplitudes and leading-color single-trace amplitudes
\be
A^\Twotwo  (1,2; ~3,4; ~5,6) =
\frac{1}{2} \left[ 
S_{12}^\Twoone + 
S_{34}^\Twoone + 
S_{56}^\Twoone \right]
- \frac{1}{6} \sum_{i<j} S_{ij}^\Twoone
+ \sum_{\sig \in S_6/\IZ_6}  A^\Twozero (\sig)
\label{222iden}
\ee
where $A^\Elltk (1,2;~3,4;~5,6)$ 
is the coefficient of 
$N^{L-2k} \left[ 2 \Tr(12)  \Tr(34) \Tr(56) \right]$
in $\cA^{(L)}_6$,
and $S_{ij}=S_{ji}$ 
are the symmetrized 2;4 double-trace amplitudes
\be
S^\Elltkp_{ij} \equiv 
A^\Elltkp(i,j; l,p,q,r) +
A^\Elltkp(i,j; l,r,p,q) +
A^\Elltkp(i,j; l,q,r,p)
\label{sym24}
\ee
where $\{ i,j \} \cup \{ l,p,q,r \} =\{1,2,3,4,5,6\}$.
By virtue of \eqn{reflection} and cyclicity of the trace, 
$S_{ij}$ is completely symmetric in $l$, $p$, $q$, and $r$,
which can therefore be suppressed.
By virtue of eq.~(\ref{eq:6pt2Lp}~e),
the fifteen $S_{ij}$ are not independent but obey the
five independent relations
\be
S_{ij}^\Twoone + 
S_{il}^\Twoone + 
S_{jl}^\Twoone 
=
S_{pq}^\Twoone + 
S_{pr}^\Twoone + 
S_{qr}^\Twoone 
\label{sym24rln}
\ee
where $\{ i,j,l,p,q,r \} $ is any permutation of $\{ 1,2,3,4,5,6 \} $.

Because the irreducible representation $N \repi{3;3}{\twooneoneoneone}$
does not appear in the two-loop null space (\ref{eq:6pt2Lp}),
it is not possible to solve for the 3;3 double-trace amplitude 
$A^\Twoone (1,2,3; ~4,5,6)$ in terms of other amplitudes.
Equations (\ref{eq:6pt2Lp}~d,g), however,
contain the irreducible representations
$N \repi{3;3}{\six}$ and $N \repi{3;3}{\fourtwo}$,
and are therefore sufficient to solve for the 
10 symmetrized 3;3 double-trace amplitudes (\ref{sym33})
\be
S^\Twoone_{123} = 
~-~\frac{1}{2} \left[ 
S_{12}^\Twoone + 
S_{23}^\Twoone + 
S_{13}^\Twoone \right]
+ \frac{1}{2}  \sum_{\sig \in S_6/\IZ_6}  A^\Twozero (\sig) \,.
\label{sym33iden}
\ee
All 66 of the relations 
(\ref{SCiden}),
(\ref{222iden}),
(\ref{sym24rln}),
and 
(\ref{sym33iden}),
which follow from 
eqs.~(\ref{eq:6pt2Lp}~a,d-m),
can alternatively be derived using U(1) decoupling. 

Because the irreducible representation $N \repi{2;4}{\threeoneoneone}$
does not appear in the two-loop null space (\ref{eq:6pt2Lp}),
it is not possible to solve for the 2;4 double-trace amplitude 
$A^\Twoone (1,2; ~3,4,5,6)$ in terms of other amplitudes.
It is possible, however, to express the symmetrized 2;4 double-trace
amplitudes $S_{ij}$,
which correspond to the irreducible representations 
$N \repi{2;4}{\six}$, $N \repi{2;4}{\fiveone}$, 
and 
$N \repij{2;4}{\fourtwo}{1} + \half N \repij{2;4}{\fourtwo}{2}$,
in terms of other amplitudes, namely,
\ba
\label{sym24iden}
S^\Twoone_{12} &= &
~-~\frac{2}{3} \left[ 
S_{123}^\Twoone + 
S_{124}^\Twoone + 
S_{125}^\Twoone + 
S_{126}^\Twoone\right] 
\\
&&
~+~\frac{1}{3} \left[ 
S_{134}^\Twoone + 
S_{135}^\Twoone + 
S_{136}^\Twoone + 
S_{145}^\Twoone + 
S_{146}^\Twoone + 
S_{156}^\Twoone \right]
+ \frac{1}{3}  \sum_{\sig \in S_6/\IZ_6}  A^\Twozero (\sig) \,.
\nn
\ea
The 15 relations (\ref{sym24iden}) are equivalent to 
\eqns{sym24rln}{sym33iden},
and therefore also follow from U(1) decoupling relations.

In total, there are 80 group-theory constraints among 200 color-ordered
amplitudes at the two-loop level, 
66 of them derivable from U(1) decoupling and 14 that are not.

\vfil\break
\bigsk\noindent{{\bf Three-loop six-point null space}}

\noin
At the three-loop level, 
there are 76 null vectors, which decompose into irreducible subspaces spanned by
\begin{subequations} \label{eq:6pt3Lp}
\begin{flalign}
& N^3 \rep{6}{\six}-\nicefrac{1}{20} N^2\rep{2;4}{\six} \nn \\
&\qquad-\nicefrac{1}{   30} N\rep{2;2;2}{\six} +\nicefrac{1}{75} \rep{2;4}{\six}  \\ \nn \\
& N^2 \rep{2;4}{\six}+\twothirds N^2 \rep{3;3}{\six} \nn \\
& \qquad-\nicefrac{1}{   3} N \rep{2;2;2}{\six} +\nicefrac{1}{30} \rep{2;4}{\six}  \\ \nn \\
& N^2 \repij{2;4}{\fourtwo}{1}+\half N^2 \repij{2;4}{\fourtwo}{2} \nn \\
& \qquad +\third N^2 \repi{3;3}{\fourtwo} -\nicefrac{1}{   8} N \repi{2;2;2}{\fourtwo} &i=1,\cdots,9 \\ \nn \\
& N \rep{6}{\six}-\nicefrac{1}{20} \rep{2;4}{\six}  \\
& N \repij{6}{\fourtwo}{1} +\fourth N \repij{6}{\fourtwo}{2} -\nicefrac{1}{8} \repij{2;4}{\fourtwo}{1}& i=1,\cdots,9  \\
& N \repi{6}{\threetwoone}-\fourth \repi{2;4}{\threetwoone} & i=1,\cdots,16  \\
& N \repij{6}{\twotwotwo}{1}+N \repij{6}{\twotwotwo}{2} + \half \repi{2;4}{\twotwotwo}& i=1,\cdots,5  \\
& N \repi{6}{\twooneoneoneone}-\nicefrac{1}{6} \repi{3;3}{\twooneoneoneone} & i=1, \cdots,5  \\
& N \repi{2;2;2}{\twotwotwo}-2 \repi{2;4}{\twotwotwo}-\third \repi{3;3}{\twotwotwo}& i=1,\cdots,5  \\
& \rep{2;4}{\six}+\twothirds \rep{3;3}{\six}  \\
& \repi{2;4}{\fiveone} & i=1,\cdots,5  \\
& \repij{2;4}{\fourtwo}{1}+\half \repij{2;4}{\fourtwo}{2} & i=1,\cdots,9  \\
& \repi{3;3}{\fourtwo} & i=1,\cdots,9
\end{flalign}
\end{subequations}
As we will see, eqs.~(\ref{eq:6pt3Lp}~d-m) 
continue to hold at all higher odd-loop levels,
and the three-loop constraints derived from them
also apply at all higher odd-loop levels.
Equations (\ref{eq:6pt3Lp}~d-m) are identical to the
one-loop null vectors (\ref{eq:6pt1Lp}~a-j), 
except for eq.~(\ref{eq:6pt3Lp}~i)
which contains an extra triple-trace term   
$N \repi{2;2;2}{\twotwotwo}$
relative to eq.~(\ref{eq:6pt1Lp}~i).
Equations (\ref{eq:6pt3Lp}~d,e,f,g,k,l) 
allow one to solve for  the 45 most-subleading-color 2;4 double-trace amplitudes
\be
A^\Teptep(1,2; ~ 3,4,5,6 ) 
= \sum_{\sig \in COP \{2,1\} \{3,4,5,6\} } A^\Tepte(\sig)
\label{threeloopUfirst}
\ee
but the expression for the 20 most-subleading-color 
3;3 double-trace amplitudes at three-
and higher loops differs from the one-loop expression (\ref{sixptDouble})
by the addition of triple-trace terms
\ba
A^\Teptep(1,2, 3; ~4,5,6 ) 
&=& ~(-1)~ \sum_{\sig \in COP \{3,2,1\} \{4,5,6\} } A^\Tepte(\sig) \nn\\
&&+ {1 \over 2} 
\sum_{ \sig \in S_3 }  (-1)^\sig  
A^\Tepte(\sig(1),4 ; ~\sig(2), 5; ~ \sig(3),6) \,.
\label{threeloopnew}
\ea
The extra terms  in \eqn{threeloopnew} cancel from the 
symmetrized 3;3 double-trace amplitudes (\ref{sym33}),
which are thus given by  
\be
S_{123}^\Teptep 
= ~(-1)~ \sum_{\sig \in COP \{3,2,1\} } A^\Tepte(\sig)
\label{threeloopUsecond}
\ee
and have the same form as \eqn{sixptUthird}.
Both \eqns{threeloopUfirst}{threeloopUsecond} 
can be derived using U(1) decoupling
but \eqn{threeloopnew} cannot.
There are 11 additional relations that follow from
eqs.~(\ref{eq:6pt3Lp}~a,b,c)
that also cannot be derived using U(1) decoupling,
yielding altogether 76 group-theory constraints 
on three-loop color-ordered amplitudes.

\bigsk\noindent{{\bf Even-loop six-point null space} } ($L \ge 4$) 

\noin
We find 76 null vectors at the four-loop level, 
which can be decomposed into  irreducible subspaces spanned by
\begin{subequations} \label{eq:6pt4Lp}
\begin{flalign}
&N^4 \rep{6}{\six}-\nicefrac{9}{20} N^3\rep{2;4}{\six} \nn \\
&\qquad -\nicefrac{4}{15} N^3 \rep{3;3}{\six} +\nicefrac{1}{   10} N^2\rep{2;2;2}{\six} \\ \nn \\
&N^3 \repij{2;4}{\fourtwo}{1}+\half N^3 \repij{2;4}{\fourtwo}{2}+\third N^3 \repi{3;3}{\fourtwo} \nn \\
&\qquad -\nicefrac{1}{  8} N^2 \repi{2;2;2}{\fourtwo}- N^2 \repij{6}{\fourtwo}{1}-\fourth N^2 \repij{6}{\fourtwo}{2} \nn \\
&\qquad +\fourth N\repij{2;4}{\fourtwo}{1}+\nicefrac{1}{16} N \repij{2;4}{\fourtwo}{2}-\nicefrac{1}{4} \repij{6}{\fourtwo}{1} & i=1,\cdots,9 \\ \nn \\
&N^2 \rep{6}{\six}-\nicefrac{1}{20} N\rep{2;4}{\six}-\nicefrac{1}{  30} \rep{2;2;2}{\six}  \\
&N\rep{2;4}{\six}+\twothirds N\rep{3;3}{\six} -\nicefrac{1}{   3} \rep{2;2;2}{\six}  \\
&N \repi{2;4}{\fiveone} & i=1,\cdots,5  \\
&N\repij{2;4}{\fourtwo}{1}+\half N \repij{2;4}{\fourtwo}{2}-\nicefrac{1}{   4} \repi{2;2;2}{\fourtwo}& i=1,\cdots,9  \\
&N \repi{3;3}{\fourtwo} +\nicefrac{3}{   8} \repi{2;2;2}{\fourtwo}& i=1,\cdots,9   \\
&\repi{2;2;2}{\twotwotwo}& i=1,\cdots,5  \\
&\rep{6}{\six}  \\
&\repij{6}{\fourtwo}{1} +\fourth \repij{6}{\fourtwo}{2} & i=1,\cdots,9   \\
&\repi{6}{\threetwoone} & i=1,\cdots,16  \\
&\repij{6}{\twotwotwo}{1}+\repij{6}{\twotwotwo}{2} & i=1,\cdots,5  \\
&\repi{6}{\twooneoneoneone} & i=1,\cdots,5
\end{flalign}
\end{subequations}
Here eqs.~(\ref{eq:6pt4Lp}~c-m) are identical to the two-loop
null vectors (\ref{eq:6pt2Lp}~a,d-m),
and this pattern persists for all higher even-loop orders.
Hence the 66 relations derived from these equations hold for all
even-loop orders 
\be
\label{fourloopcon1}
A^\Tete (1, \setal, 6, \setbe )
=  (-1)^{n_\beta} \sum_{\sig \in OP \setal \setbeT } A^\Tete (1,\sig, 6) \,,
\ee
\ba
A^\Tete  (1,2; ~3,4; ~5,6) &=& 
\frac{1}{2} \left[ 
S_{12}^\Tetemo + 
S_{34}^\Tetemo + 
S_{56}^\Tetemo \right]
\nn\\
&& - \frac{1}{6} \sum_{i<j} S_{ij}^\Tetemo
+ \sum_{\sig \in S_6/\IZ_6}  A^\Tetemt (\sig) \,,
\ea
\be
S^\Tetemo_{123} =
~-~\frac{1}{2} \left[ 
S_{12}^\Tetemo + 
S_{23}^\Tetemo + 
S_{13}^\Tetemo \right]
+ \frac{1}{2}  \sum_{\sig \in S_6/\IZ_6}  A^\Tetemt (\sig) \,,
\ee
\be
S_{ij}^\Tetemo + 
S_{il}^\Tetemo + 
S_{jl}^\Tetemo 
=
S_{pq}^\Tetemo + 
S_{pr}^\Tetemo + 
S_{qr}^\Tetemo  \,.
\label{fourloopcon2}
\ee
Equations (\ref{fourloopcon1})-(\ref{fourloopcon2})
can be derived using U(1) decoupling (for all $\ell$).
The 14 additional two-loop relations derived from 
eqs.~(\ref{eq:6pt2Lp}~b,c) 
are replaced at four- (and higher-) loops by 10 relations that follow from 
eqs.~(\ref{eq:6pt4Lp}~a,b),
and that cannot be derived using U(1) decoupling.
Altogether there are 76 constraints on even-loop
color-ordered amplitudes for $L \ge 4$.

\bigsk\noindent{{\bf Odd-loop six-point null space }} ($L \ge 5$) 

\noin
We find 76 null vectors at the five-loop level, 
which can be decomposed into the $S_6$ representations,
\begin{subequations} \label{eq:6pt5Lp}
\begin{flalign}
&N^5 \rep{6}{\six}-\nicefrac{9}{20} N^4 \rep{2;4}{\six} -\nicefrac{4}{15} N^4 \rep{3;3}{\six} \nn \\
&\qquad  +\nicefrac{1}{   10} N^3 \rep{2;2;2}{\six} + \nicefrac{3}{4} N^3 \rep{6}{\six} - \nicefrac{3}{80} N^2 \rep{2;4}{\six} \nn \\
&\qquad - \nicefrac{1}{   40} N \rep{2;2;2}{\six} + \nicefrac{1}{100} \rep{2;4}{\six} \\ \nn \\
&N^3 \rep{6}{\six}-\nicefrac{9}{20} N^2 \rep{2;4}{\six} \nn \\
&\qquad -\nicefrac{4}{15} N^2 \rep{3;3}{\six} +\nicefrac{1}{   10} N \rep{2;2;2}{\six} \\ \nn \\
&N^2 \repij{2;4}{\fourtwo}{1}+\half N^2 \repij{2;4}{\fourtwo}{2}+\third N^2 \repi{3;3}{\fourtwo} \nn \\
&\qquad -\nicefrac{1}{  8} N \repi{2;2;2}{\fourtwo} & i=1,\cdots,9 \\ \nn \\
&N \rep{6}{\six}-\nicefrac{1}{20} \rep{2;4}{\six}  \\
&N \repij{6}{\fourtwo}{1} +\fourth N \repij{6}{\fourtwo}{2} -\nicefrac{1}{8} \repij{2;4}{\fourtwo}{1}& i=1,\cdots,9  \\
&N \repi{6}{\threetwoone}-\fourth \repi{2;4}{\threetwoone} & i=1,\cdots,16  \\
&N \repij{6}{\twotwotwo}{1}+N \repij{6}{\twotwotwo}{2} + \half \repi{2;4}{\twotwotwo}& i=1,\cdots,5  \\
&N \repi{6}{\twooneoneoneone}-\nicefrac{1}{6} \repi{3;3}{\twooneoneoneone} & i=1, \cdots,5  \\
&N \repi{2;2;2}{\twotwotwo}-2 \repi{2;4}{\twotwotwo}-\third \repi{3;3}{\twotwotwo}& i=1,\cdots,5  \\
&\rep{2;4}{\six}+\twothirds \rep{3;3}{\six}  \\
&\repi{2;4}{\fiveone} & i=1,\cdots,5  \\
&\repij{2;4}{\fourtwo}{1}+\half \repij{2;4}{\fourtwo}{2} & i=1,\cdots,9  \\
&\repi{3;3}{\fourtwo} & i=1,\cdots,9
\end{flalign}
\end{subequations}

The five-loop null vectors (\ref{eq:6pt5Lp}~d-m) 
are precisely the same as the three-loop
null vectors (\ref{eq:6pt3Lp}~d-m),
and hence give rise to the same 65 constraints
(\ref{threeloopUfirst}) and (\ref{threeloopnew}).
Of the 11 additional three-loop relations derived from 
eqs.~(\ref{eq:6pt3Lp}~a,b,c),
10 survive, but one is replaced by a relation that follows from 
eq.~(\ref{eq:6pt5Lp}~a).
None of the additional relations follow from U(1) decoupling.
Altogether there are 76 group-theory constraints on odd-level 
color-ordered amplitudes for $L \ge 5$.

After five loops, the pattern of null vectors begins to repeat.
We find that the six-loop null space is equivalent to 
the four-loop null space, and 
the seven-loop null space is equivalent to 
the five-loop null space.
By examining the structure of  \eqn{contract}, 
one can then prove that 
all even-loop null spaces with $L \ge 4$ are equivalent to the
four-loop null space, and that 
all odd-loop null spaces with $L \ge 5$ are equivalent to the
five-loop null space.
Hence the group theory constraints at four and five loops are 
repeated at all higher even- and odd-loop levels respectively.

Finally, a word about how we determined the coefficients
in the preceding equations.
As we have seen,  
a given irreducible representation of $S_6$ can appear multiple times 
in the induced representations given in 
eqs.~(\ref{sixdecomp}), (\ref{twofourdecomp}), 
(\ref{threethreedecomp}), and (\ref{twotwotwodecomp}); 
e.g., the $\twotwotwo$ representation
appears twice in the single-trace representation, and once
in each of the multiple-trace representations.
Null vectors belong to certain specific linear combinations
of these irreducible representations. 
To determine these linear combinations,
we construct the most general linear combination of a given
tableau $\tau$ that can appear in the $L$-loop trace space.
For example, the most general five-dimensional representation
$\twotwotwo$ appearing in the two-loop trace basis  is
\ba
u_i &=& \kappa_1  N^2 \repij{6}{\twotwotwo}{1} 
+\kappa_2 N^2 \repij{6}{\twotwotwo}{2} 
+\kappa_3 N \repi{2;4}{\twotwotwo}
+\kappa_4 N \repi{3;3}{\twotwotwo}
\nn\\
&&
+\kappa_5 \repi{2;2;2}{\twotwotwo}
+\kappa_6  \repij{6}{\twotwotwo}{1} 
+\kappa_7 \repij{6}{\twotwotwo}{2} \,.
\label{ansatz}
\ea
We expand $u_i$ in the two-loop trace basis
\be 
u_i = \sum_{\la=1}^{200} t^\Two_\la  u_{\la i} \,.
\ee
It is only necessary to carry out this 
procedure for a single element, e.g., $i=1$,
because if $u_1$ belongs to the null space, so does the 
rest of the irreducible representation.
We then act on $u_1$ with $M^\Two$,
calculated as described earlier in this section,
and set the result to zero to determine 
the conditions that must be satisfied by the coefficients  $\kappa$.
In this case we find
\be
2 \kappa_1= 2 \kappa_2 = 20 \kappa_3 = -30\kappa_4 = 15(\kappa_7-\kappa_6)
\ee
allowing for three independent solutions 
which are precisely those
occurring in the two-loop null space, 
viz. eqs.~(\ref{eq:6pt2Lp}~c,h,l).
Although we used a specific choice for the  trace basis to determine the coefficients,
the result is independent of this choice.

Another consistency check that we found useful is the following.
If we take any $L$-loop null vector, drop the $N$-independent part,
and divide the rest by $N$, the result must necessarily belong to
the $(L-1)$-loop null space.   This follows from \eqn{recursion}
if we set $G$ in \eqn{defG} equal to $N$ times the unit matrix.

\section{Kleiss-Kuijf relations at higher loops?} 
\setcounter{equation}{0}
\label{sec-KK}

Throughout this paper, we have encountered the tree-level Kleiss-Kuijf (KK)
relations \cite{Kleiss:1988ne,DelDuca:1999rs} 
\be
A^\Zerozero (1, \setal, n, \setbe )
= (-1)^{n_\beta} \sum_{\sig \in OP \setal \setbeT } 
A^\Zerozero (1,\sig, n) \,.
\label{KKtree}
\ee
This set of $\half (n-3) (n-2)!$ relations
follows from U(1) decoupling for $n \le 6$,
and U(1) decoupling arguments also apply 
to the most-subleading-color single-trace amplitudes 
at all even-loop orders
\be
A^\Tete (1, \setal, n, \setbe )
= (-1)^{n_\beta} \sum_{\sig \in OP \setal \setbeT } 
A^\Tete (1,\sig, n),  \qquad n  = 4, 5, 6 
\label{KKhigher} 
\ee
as we have seen in 
eqns.~(\ref{fourptUfirst}), (\ref{fiveptUthird}),
and (\ref{fourloopcon1}).

For $n \ge 7$, however, the tree-level KK relations 
cannot be derived from U(1) decoupling \cite{Kleiss:1988ne},
and so the question naturally arises whether \eqn{KKhigher} holds for $n \ge 7$,
i.e., whether the KK relations apply at loop level.
Feng et. al \cite{Feng:2011fja} found that for 
$n=7$, the relations (\ref{KKhigher})
hold at two loops, but were not able to establish them for 
$n=8$.

We have confirmed that the 240 group-theory relations obeyed by the
two-loop subleading-color single-trace amplitudes are 
identical to those obeyed by the tree-level amplitudes for $n=7$. 
However, we found that,
while the tree-level eight-point amplitudes obey
the 1800 independent relations given by \eqn{KKtree},
the two-loop subleading-color single-trace eight-point amplitudes 
satisfy only 1786 group-theory relations among themselves, 
14 shy of the number necessary to establish \eqn{KKhigher} for $\ell=1$.
Hence the relations (\ref{KKhigher}) are {\it not} valid for $n = 8$ and $\ell=1$
(or at least cannot be derived from group theory alone)\footnote{We also
checked that the group-theory relations for two-loop subleading-color 
single-trace nine-point amplitudes are 
not sufficient to establish \eqn{KKhigher}.}.

The eight-point single-trace representation 
decomposes into the following irreducible representations of $S_8$
\begin{align}
{R^{(\text{ind})}_{8}\atop 2520}{~=~\atop~=~}&{\eight \atop 1} {~\oplus~\atop~+~}{3~\sixtwo \atop 3\cdot20}{~\oplus~\atop~+~}{\fivethree \atop 28}{~\oplus~\atop~+~}{4~\fivetwoone \atop 4\cdot64}{~\oplus~\atop~+~}{2~\fiveoneoneone \atop 2\cdot35}{~\oplus~\atop~+~}{3~\fourfour \atop 3\cdot14}{~\oplus~\atop~+~}{3~\fourthreeone \atop 3\cdot70} \nn \\[3mm]
&{~\oplus~\atop~+~}{7~ \fourtwotwo \atop 7\cdot56}
{~\oplus~\atop~+~}{4~\fourtwooneone \atop 4\cdot90}{~\oplus~\atop~+~}{4~\fouroneoneoneone \atop 4\cdot35}{~\oplus~\atop~+~}{\threethreetwo \atop 42}{~\oplus~\atop~+~}{5~\threethreeoneone \atop 5\cdot56}{~\oplus~\atop~+~}{4~\threetwotwoone \atop 4\cdot70}{~\oplus~\atop~+~}{4~\threetwooneoneone \atop 4\cdot64} \nn \\[3mm]
&{~\oplus~\atop~+~}{\threeoneoneoneoneone \atop 21}{~\oplus~\atop~+~}{3~\twotwotwotwo \atop 3\cdot 14}{~\oplus~\atop~+~}{2~\twotwooneoneoneone \atop 2\cdot20}
\end{align}
which contains 6 copies of 14-dimensional representations.
We conclude that one of these representations contained in the tree-level eight-point
null space must be absent from the two-loop eight-point null space,
although we have not tried to ascertain which.

In conclusion, for $n \ge 8$,  
the KK relations do not extend beyond tree level (or at least
cannot be established using group theory alone).

\section{Concluding remarks}
\setcounter{equation}{0}
\label{sec-concl}

In this paper, we have extended the iterative approach of 
refs.~\cite{Naculich:2011ep,Edison:2011ta}
to obtain the complete set of relations 
(at all loop orders)
obeyed by color-ordered six-point amplitudes 
derivable from SU($N$) group theory alone.
We have also shown how group-theory relations among $n$-point
amplitudes can be decomposed into subsets associated with
irreducible representations of the symmetric group $S_n$
acting on the external legs.    
We used this to present the $L$-loop relations 
among four-, five-, and six-point amplitudes 
in a compact way by listing the irreducible subspaces
of the $L$-loop, $n$-point null space.

At tree level, our six-point results reproduce the 
36 Kleiss-Kuijf relations (\ref{sixptUfirst}), 
which follow from U(1) decoupling,
and at one loop, we reproduce the 65 relations
(\ref{sixptUsecond}, \ref{sixptDouble})
expressing the double-trace amplitudes in terms
of single-trace amplitudes originally found in 
refs.~\cite{Bern:1990ux,Bern:1994zx}.
At two loops, we obtained 80 relations, 
of which 66 are derivable from U(1) decoupling and 14 are not. 
The 66 two-loop relations include 36 Kleiss-Kuijf relations
among the subleading-color single-trace amplitudes
(\ref{SCiden}),
15 relations expressing the triple-trace amplitudes
(\ref{222iden}),
and 15 relations among double-trace and leading-color single-trace amplitudes
(\ref{sym24rln}, \ref{sym33iden}).
At three loops, we obtained 76 relations,
of which 65 can be used to express the most-subleading-color 
double-trace amplitudes in terms
of single- and triple-trace amplitudes.
Of these 65 three-loop relations, 
55 can be derived from U(1) decoupling 
(\ref{threeloopUfirst}, \ref{threeloopUsecond})
and have a form analogous to the one-loop relations.
The other three-loop relations 
(\ref{threeloopnew})
differ from the one-loop relations (\ref{sixptDouble})
by the addition of triple-trace amplitudes.
At all higher loops, there are 76 group-theory relations
among the color-ordered amplitudes, 
and the pattern of relations begins to repeat after
five loops.   Further details may be found in sec.~\ref{sec-six}.

We should mention one final caveat.
In our iterative approach, 
we made the assumption that one can construct 
the complete $(L+1)$-loop color space by attaching rungs 
between the external legs of $L$-loop color factors in all possible ways.
There are no known counterexamples to this assumption,
but neither has it been proved.
If this assumption were incorrect,
then some of the $L$-loop color spaces
would be larger than we have surmised, 
and the complementary null spaces smaller.  
Consequently, some of the irreducible subspaces of $S_n$
would be absent from the null spaces, 
and there would be fewer group-theory relations than we have stated.
In no case, however, can there be any additional null vectors 
or new group-theory relations among color-ordered amplitudes.

\section*{Acknowledgments}

It is a pleasure to thank W. Barker, L. Dixon, and T. Pietraho
for helpful discussions and remarks.    
We are especially grateful to C. Boucher-Veronneau for her 
involvement in the early stages of this project.

\vfil\break

\appendix 
\section{Decomposition of trace representations} 
\setcounter{equation}{0}
\label{app-char}

To find the decomposition 
into irreducible representations of $S_n$
of a representation invariant under a subgroup
$B\subset S_n$ 
we note that the number of times a given irreducible representation
$\tau$ appears in the decomposition
is given by \cite{Schensted}
\be 
\label{eq:deltadef} 
\gamma^{\tau}
= \frac{1}{n_B}\sum_{\sigma\in B} \chi^{\tau}(\sigma), 
\ee 
where
$\chi^{\tau}(\sigma)$ is the character of $\sigma$ in the irreducible
representation $\tau$ of $S_n$,
and $n_B$ is the order of $B$.

For $n$ even, the single-trace basis 
\be
\label{eq:singletrace}
\Tr(123\cdots n) +(-1)^n \Tr(n\cdots321)
\ee
 is invariant under the dihedral subgroup $D_n \subset S_n$ of order $2n$,
generated by cyclic and reflection permutations.  
Taking, for example, the four-point case, the elements of $D_4$ are
\be
\{1,~~
(24),~~(12)(34),~~
(1234),~~
(13),~~
(13)(24),~~
(1432),~~
(14)(23)\}.
\ee
Since the characters only depend on the conjugacy class of $\sigma$,
we identify each element's conjugacy class, 
which is determined by the lengths and numbers of cycles
\begin{align}
&1&&(24)&&(12)(34)&&(1234)&&(13)&&(13)(24)&&(1432)&&(14)(23) \nn \\
&\downarrow&&\downarrow&&\downarrow&&\downarrow&&\downarrow&&\downarrow&&\downarrow&&\downarrow \nn \\
&[1^4]&&[2\ 1^2]&&[2^2]&&[4]&&[2\ 1^2]&&[2^2]&&[4]&&[2^2]
\end{align}
We then use \eqn{eq:deltadef} to compute 
\begin{align}
\gamma^{\tau}&=\frac{1}{8}\left[\chi^\tau ([1^4])+2\chi^{\tau}([2\ 1^2])+3\chi^\tau([2^2])+2\chi^\tau([4])\right],
\end{align}
yielding
\begin{equation}
\gamma^{\four} = 1,
\qquad \gamma^{\threeone}=0,
 \qquad \gamma^{\twotwo}=1,
 \qquad \gamma^{\twooneone}=0,
 \qquad \gamma^{\oneoneoneone}=0.
\end{equation}
Thus for the four-point case, the single-trace basis decomposes into
\be
R^\ind_4 = \six \ \oplus \ \twotwo \ .
\ee

For $n$ odd, however, the trace basis (\ref{eq:singletrace})
is only invariant 
up to a sign under the dihedral subgroup.
$D_n$ contains two different types of permutations: $n$ cyclic permutations $\sigma_{\cyc}$ and $n$ reversal permutations $\sigma_{\rev}$.
These act differently on the antisymmetric basis element 
\begin{align}
\sigma_{\cyc}[\Tr(123\cdots n)-\Tr(n\cdots 321)] &= \Tr(123\cdots n)-\Tr(n\cdots 321) \nn \\
\sigma_{\rev}[\Tr(123\cdots n)-\Tr(n\cdots 321)]&=-[\Tr(123\cdots n)-\Tr(n\cdots 321)].
\end{align}
To deal with this, we observe that $D_n$ has 
(at least) two one-dimensional representations, the trivial representation with
\begin{equation}
\chi_{D_n}^{T}(\sigma)=1
\end{equation}
and the non-trivial one-dimensional representation with
\be
\chi_{D_n}^{NT}(\sigma_{\cyc})=1, \qquad
\chi_{D_n}^{NT}(\sigma_{\rev})=-1.
\ee
Hence, when $n$ is odd, 
the trace basis elements are invariant under the action of $\chi_{D_n}^{NT}(\sigma)\sigma$:
\begin{equation}
\chi_{D_n}^{NT}(\sigma)\sigma~[\Tr(123\cdots n)-\Tr(n\cdots 321)] = \Tr(123\cdots n)-\Tr(n\cdots 321),
\end{equation}
and we can use this to modify  \eqref{eq:deltadef}
\be
\gamma^{\tau} = \frac{1}{n_B}\sum_{\sigma\in B} \chi^{R}_{D_n}(\sigma)\chi^{\tau}(\sigma), \qquad R= 
\begin{cases} T, & \hbox{ for  $n$ even,} \\
              NT, & \hbox{ for  $n$ odd.}
\end{cases}
\ee

Multiple-trace representations can be treated in a similar fashion.

\section{Five-point decomposition in  an explicit trace basis} 
\setcounter{equation}{0}
\label{app-five}

In sec.~\ref{sec-five}, we presented the five-point null spaces 
in terms of irreducible $S_5$  subspaces in a form independent of the
choice of basis.  
In this appendix, we specify an explicit basis in order 
to facilitate comparison of the results of this paper with those 
of ref.~\cite{Edison:2011ta}.

The single-trace basis we use is 
\ba
T_1     &=& \left[\Tr (12345) - \Tr(15432)\right],\qquad\qquad
T_7       = \left[\Tr (12543) - \Tr(13452)\right],\nonumber\\
T_2     &=& \left[\Tr (14325) - \Tr(15234)\right],\qquad\qquad
T_8       = \left[\Tr (14523) - \Tr(13254)\right],\nonumber\\
T_3     &=& \left[\Tr (13425) - \Tr(15243)\right],\qquad\qquad
T_9       = \left[\Tr (13524) - \Tr(14253)\right],\nonumber\\
T_4     &=& \left[\Tr (12435) - \Tr(15342)\right],\qquad\qquad
T_{10}   =  \left[\Tr (12534) - \Tr(14352)\right],\nonumber\\
T_5     &=& \left[\Tr (14235) - \Tr(15324)\right],\qquad\qquad
T_{11}   =  \left[\Tr (14532) - \Tr(12354)\right],\nonumber\\
T_6     &=& \left[\Tr (13245) - \Tr(15423)\right],\qquad\qquad
T_{12}   =  \left[\Tr (13542) - \Tr(12453)\right].
\label{singletrace}
\ea
The vectors (\ref{fivepointsingletracedecomp})
that span the irreducible representations of the single-trace representation
$R^\ind_5$ can be expanded in this basis as
\be
\repij{5}{\tau}{j} 
=
\sum_{\la=1}^{12} T_\la \replij{5}{\tau}{j} 
\ee
where
\be
\replij{5}{\threeoneone}{1} 
= 
\tenth
 \left(
                 \begin{smallmatrix}
                  2 & 1 & 1 & 2 & 1 & 2 \\
                  2 & 1 & -2\,\ & 2 & -1\,\ & -1\,\ \\
                  1 & 2 & 2 & 2 & -1\,\ & -1\,\ \\
                  1 & 2 & -1\,\ & 2 & 2 & 1 \\
                  2 & 2 & -1\,\ & 1 & -2\,\ & 1 \\
                  2 & 2 & 1 & 1 & 1 & -2\,\ \\
                  -1\,\ & -1\,\ & -2\,\ & 2 & 1 & 2 \\
                  2 & -1\,\ & -2\,\ & -1\,\ & -1\,\ & 2 \\
                  -1\,\ & 2 & 2 & -1\,\ & 2 & -1\,\ \\
                  -1\,\ & -1\,\ & 2 & 2 & 2 & 1 \\
                  2 & -1\,\ & -1\,\ & -1\,\ & -2\,\ & -2\,\ \\
                  -1\,\ & 2 & 1 & -1\,\ & -2\,\ & -2\,\ \\
                 \end{smallmatrix}
                 \right) \,,
\quad
\replij{5}{\threeoneone}{2} 
= 
\tenth
 \left(
                 \begin{smallmatrix}
                  -1\,\ & 2 & 2 & -1\,\ & 2 & -1\,\ \\
                  -1\,\ & 2 & 1 & -1\,\ & -2\,\ & -2\,\ \\
                  2 & -1\,\ & -1\,\ & -1\,\ & -2\,\ & -2\,\ \\
                  2 & -1\,\ & -2\,\ & -1\,\ & -1\,\ & 2 \\
                  -1\,\ & -1\,\ & -2\,\ & 2 & 1 & 2 \\
                  -1\,\ & -1\,\ & 2 & 2 & 2 & 1 \\
                  -2\,\ & -2\,\ & 1 & -1\,\ & 2 & -1\,\ \\
                  -1\,\ & -2\,\ & 1 & -2\,\ & -2\,\ & -1\,\ \\
                  -2\,\ & -1\,\ & -1\,\ & -2\,\ & -1\,\ & -2\,\ \\
                  -2\,\ & -2\,\ & -1\,\ & -1\,\ & -1\,\ & 2 \\
                  -1\,\ & -2\,\ & -2\,\ & -2\,\ & 1 & 1 \\
                  -2\,\ & -1\,\ & 2 & -2\,\ & 1 & 1 \\
                 \end{smallmatrix}
                 \right) \,.
\label{explicitsixdimrep}
\ee
The linear combination of these representations that spans 
the tree-level null space (\ref{eq:5ptTree}) is
\be
3 \replij{5}{\threeoneone}{1} + \replij{5}{\threeoneone}{2} 
=
\half
\left(
                 \begin{smallmatrix}
                  1 & 1 & 1 & 1 & 1 & 1 \\
                  1 & 1 & -1~\, & 1 & -1~\, & -1~\, \\
                  1 & 1 & 1 & 1 & -1~\, & -1~\, \\
                  1 & 1 & -1~\, & 1 & 1 & 1 \\
                  1 & 1 & -1~\, & 1 & -1~\, & 1 \\
                  1 & 1 & 1 & 1 & 1 & -1~\, \\
                  -1~\, & -1~\, & -1~\, & 1 & 1 & 1 \\
                  1 & -1~\, & -1~\, & -1~\, & -1~\, & 1 \\
                  -1~\, & 1 & 1 & -1~\, & 1 & -1~\, \\
                  -1~\, & -1~\, & 1 & 1 & 1 & 1 \\
                  1 & -1~\, & -1~\, & -1~\, & -1~\, & -1~\, \\
                  -1~\, & 1 & 1 & -1~\, & -1~\, & -1~\, \\
                 \end{smallmatrix}
                 \right) \,.
\label{explicittreenull}
\ee
These are linear combinations of the six column vectors of $x^\Zero$ 
given in eq.~(3.11) of ref.~\cite{Edison:2011ta},
and so span the same space.

The double-trace basis we use is 
\ba
T_{13} &=&  \Tr (12) \left[ \Tr(345) - \Tr(543) \right], \qquad\qquad
T_{18}   =  \Tr (13) \left[ \Tr(245) - \Tr(542) \right], \nonumber\\
T_{14} &=&  \Tr (23) \left[ \Tr(451) - \Tr(154) \right], \qquad\qquad
T_{19}   =  \Tr (24) \left[ \Tr(351) - \Tr(153) \right], \nonumber\\
T_{15} &=&  \Tr (34) \left[ \Tr(512) - \Tr(215) \right], \qquad\qquad
T_{20}   =  \Tr (35) \left[ \Tr(412) - \Tr(214) \right], \nonumber\\
T_{16} &=&  \Tr (45) \left[ \Tr(123) - \Tr(321) \right], \qquad\qquad
T_{21}   =  \Tr (41) \left[ \Tr(523) - \Tr(325) \right], \nonumber\\
T_{17} &=&  \Tr (51) \left[ \Tr(234) - \Tr(432) \right], \qquad\qquad
T_{22}   =  \Tr (52) \left[ \Tr(134) - \Tr(431) \right].
\label{doubletrace}
\ea
The vectors (\ref{fivepointdoubletracedecomp})
that span the irreducible representations of the double-trace representation
$R^\ind_{2;3}$ can be expanded in this basis as
\be
\repij{2;3}{\tau}{j} 
=\sum_{\la=13}^{22} T_\la \repli{2;3}{\tau} 
\ee
where
\be
\repli{2;3}{\threeoneone}
 =
\nicefrac{1}{5} 
 \left(
                 \begin{smallmatrix}
                  1 & -1\,~ & 1 & 0 & 0 & 0 \\
                  3 & 1 & -1\,~ & 1 & -1\,~ & 0 \\
                  1 & 1 & 0 & 3 & 1 & 1 \\
                  0 & -1\,~ & -1\,~ & 1 & 1 & 3 \\
                  0 & 0 & 1 & 0 & -1\,~ & 1 \\
                  1 & 0 & 0 & -1\,~ & 1 & 0 \\
                  1 & 3 & 1 & 1 & 0 & -1\,~ \\
                  -1\,~ & 0 & 1 & 1 & 3 & 1 \\
                  0 & 1 & 0 & -1\,~ & 0 & 1 \\
                  -1\,~ & 1 & 3 & 0 & 1 & -1\,~ \\
                 \end{smallmatrix}
                 \right) \,,
\qquad
\repli{2;3}{\twooneoneone}
= 
\nicefrac{2}{5} 
                 \left(\begin{smallmatrix}
                  1 & 0 & 0 & 0 \\
                  -1\,~ & -1\,~ & 0 & 0 \\
                  0 & 1 & 1 & 0 \\
                  0 & 0 & -1\,~ & -1\,~ \\
                  0 & 0 & 0 & 1 \\
                  0 & 1 & 0 & 0 \\
                  1 & 0 & -1\,~ & 0 \\
                  0 & -1\,~ & 0 & 1 \\
                  0 & 0 & 1 & 0 \\
                  -1\,~ & 0 & 0 & -1\,~ \\
                 \end{smallmatrix}
                 \right) \,.
\ee
By using these explicit decompositions,
it is straightforward to show that the 
vectors spanning odd- and even-loop five-point null spaces 
(\ref{eq:5pt1Lp}) and (\ref{eq:5pt2Lp}) 
are linear combinations of the odd- and even-loop null vectors
written down in ref.~\cite{Edison:2011ta}.

\vfil\break

\end{document}